\documentclass[prl,aps,floats,twocolumn,superscriptaddress,floatfix,longbibliography]{revtex4-1}
\usepackage{amssymb,amsmath}
\usepackage{amsmath,amssymb}
\usepackage[mathscr]{euscript}
\usepackage{graphicx}
\usepackage{psfrag}
\usepackage[normalem]{ulem}
\usepackage{xcolor}
\usepackage[%
  colorlinks=true,
  urlcolor=blue,
  linkcolor=blue,
  citecolor=blue
]{hyperref}

\def\beq{\begin{equation}}
\def\eeq{\end{equation}}
\def\bea{\begin{eqnarray}}
\def\eea{\end{eqnarray}}

\definecolor{mygreen}{rgb}{0.0,0.55,0.3}

\begin{document}
 \title{Theory of Hyperuniformity at the Absorbing State Transition}
\author{Xiao Ma}
\affiliation{DAMTP, Centre for Mathematical Sciences, University of Cambridge, Wilberforce Road, Cambridge, CB3 0WA}
\author{Johannes Pausch}
\affiliation{DAMTP, Centre for Mathematical Sciences, University of Cambridge, Wilberforce Road, Cambridge, CB3 0WA}
\affiliation{Department of Mathematics, Imperial College, London SW7 2AZ}
\author{Michael E. Cates}
\affiliation{DAMTP, Centre for Mathematical Sciences, University of Cambridge, Wilberforce Road, Cambridge, CB3 0WA}

\date{\today}
\begin{abstract}
Hyperuniformity, whereby the static structure factor (or density correlator) obeys $S(q)\sim q^{\varsigma}$ with $\varsigma> 0$,  emerges at criticality in systems having multiple absorbing states, such as periodically sheared suspensions. These lie in the conserved directed percolation (C-DP) universality class, for which analytic results for $\varsigma$ are lacking. Specifically, $\varsigma$ appears inaccessible within an exact `interfacial mapping' that yields other C-DP exponents via functional renormalization group (FRG). Here, 
using Doi-Peliti field theory for interacting particles and perturbative RG about a Gaussian model, we find $\varsigma = 0^+$ and $\varsigma= 2\epsilon/9 + O(\epsilon^2)$ in dimension $d>4$ and $d=4-\epsilon$ respectively. The latter disproves a previously conjectured scaling relation for $\varsigma$. We show how hyperuniformity emerges from anticorrelation of  strongly fluctuating active and passive densities. Our calculations also yield the remaining C-DP exponents without recourse to functional RG methods.
\end{abstract} 

\maketitle

Any configuration that a system can enter, but not escape from, is called an absorbing state. For example, experiments on non-Brownian particles of number density $\rho$ supended in a fluid, subject to slow periodic shearing of fixed amplitude, show that at large $\rho$ particles collide and are randomly displaced each cycle: there is always a finite density, $\rho_A$, of `active' particles. In contrast, below a critical density, $\rho = \rho_c$, particles `randomly organize' into a non-colliding, stroboscopically static, disordered state in which all particles are passive: $\rho \equiv \rho_A+\rho_P = \rho_P$ \cite{RO1,RO2,RO3}. At 
 $\rho = \rho_c$, this passive state takes infinitely long to appear, and acquires infinitely long-range correlations as is generic at a second order phase transition. 

Remarkably, these spatial correlations are hyperuniform \cite{Torq}: avoidance of collisions requires density fluctuations to {\em vanish} at low wavenumbers $q$, rather than {\em diverge} as in equilibrium criticality. In dimensions $d = 2,3$, the static structure factor at criticality vanishes as a power law: $\langle \rho_{\bf q}\rho_{-{\bf q}}\rangle \equiv S(q)\sim q^\varsigma$ with $\varsigma > 0$ \cite{HexLev,TjhungBerthier}. (Scaling arguments then give $S(0)\sim \xi^{-\varsigma}$ at large but finite correlation length $\xi$.) These phenomena are not limited to sheared colloids, but generic for systems in which a nonconserved scalar order parameter ($\rho_A$),  is coupled locally to a conserved density ($\rho$), such that there are multiple absorbing states ($\rho_A = 0$) of different frozen density patterns. This scenario defines the Conserved Directed Percolation, or C-DP, universality class \cite{Hinrichsen, Lubeck}. 

The C-DP class includes not only random organization \cite{Sriram} but several different-looking models, including the Manna model of sandpiles \cite{Manna}, and the quenched Edwards-Wilkinson model (q-EW) of interfacial growth. The q-EW correspondence uses an exact mapping  \cite{Doussal,Janssen} in which the interfacial height is the time-integrated active particle density, $u(t) = \int_0^t\rho_A(s)ds$, while the random field evaluated at height $u$ gives $\rho_P$. (Both freeze at the depinning transition.) functional renormalization group (FRG) methods applied to q-EW \cite{qEWexponents} have allowed calculation to order $\epsilon = 4-d$ of the C-DP exponents $\beta = 1-\epsilon/9$, $\nu_\perp = \frac{1}{2}+\epsilon/12$ and $z = 2-2\epsilon/9$ \cite{Doussal,Janssen}, describing the vanishing of the order parameter ($\rho_A \sim (\rho-\rho_c)^\beta$), and the divergences of the correlation length ($\xi \sim (\rho-\rho_c)^{-\nu_\perp}$) and time ($\mathcal{T}\sim\xi ^z$).

However, the exact q-EW mapping offers no clear route to calculating $S(q) = \langle \rho_{\bf q}\rho_{-{\bf q}}\rangle$ and its hyperuniformity exponent $\varsigma$. 
In this Letter we find $\varsigma$ by instead using a perturbative RG (not FRG) for a Doi-Peliti field theory of interacting particles \cite{Tauber}. This calculation encounters an ambiguity concerning the scaling dimensions of active and passive particle field operators, which can be resolved by demanding that the critical fluctuations are indeed hyperuniform, rather than divergent. Imposing this, we also confirm the three known exponent values reported above; this gives a powerful check on our methods.
Our new result, $\varsigma = 2\epsilon/9$, disproves a conjectured scaling relation due to Hexner and Levine \cite{HexLev} which, if true, would have offered a shorter and more elegant path to finding $\varsigma$ analytically. The Hexner-Levine conjecture reads, in our notation, $\varsigma = d-2\beta/\nu_\perp = \epsilon/9+O(\epsilon^2)$. 

En route to these results, we examine our interacting particle theory at Gaussian level, applicable for $d>4$. Surprisingly, we also find a type of hyperuniformity here, albeit of a singular form that can be viewed as an exponent of $\varsigma = 0^+$. This resolves uncertainty \cite{Mari, Wilken} over whether hyperuniformity persists for long-range models and/or in high dimensions where the Gaussian theory should hold. The Gaussian theory lays bare a significant feature of C-DP (hinted at in \cite{HexLev}): the fluctuations of $\rho_A$ and $\rho_P$ are not separately hyperuniform even as those of $\rho = \rho_A+\rho_P$ become so. This requires near-perfect anticorrelation between the two particle types, which our Gaussian results expose, and our RG results further illuminate. 
Notably also, we find that the conservative noise associated with diffusion of $\rho_A$, while often considered subdominant for C-DP problems  \cite{Janssen,Wiese,Sriram}, is needed to make sense of the critical hyperuniformity found at Gaussian level.

Our calculations therefore (i) unveil the true character of the C-DP transition, with hyperuniformity emerging from cancellation of large active and passive fluctuations; (ii) compute the previously elusive hyperuniformity exponent as  $\varsigma=0^+$ for $d>4$ and $\varsigma = 2\epsilon/9$ for $d<4$; (iii) disprove the Hexner-Levine conjecture that $\varsigma = d-2\beta/\nu_\perp$; (iv) show that conservative noise is required to fully understand the C-DP class; and (v) show that FRG is not necessarily required to compute its exponents.

{\em Field Theory:}
We consider a lattice model comprising $A$ particles that hop with diffusivity $D$ and non-hopping $P$ particles. The on-site reaction $A+P\to 2A$ has rate $\kappa$, causing passive particles to awaken on encounter with active ones; the reaction $A\to P$ has rate $\mu$ so that active particles decay to passivity without collisions. Following established procedures \cite{Tauber, Biroli} we write the master equation for the model using annihilation operators $a,p$ and creation operators  $a^\dagger = \tilde a +1, p^\dagger = \tilde p +1$ for $A$ and $P$ particles respectively, where site- and time-indices are suppressed to ease notation. Via a coherent-state path integral and the continuum limit, we arrive at a Doi-Peliti action $\mathcal{A} = \int \mathbb{A} \, d^d x\, dt$ with action density \cite{Tauber,SM}
\bea
\label{eq:action}
    \mathbb{A}&= -\Tilde{a}(\partial_t-D\nabla^2)a-\Tilde{p}\partial_t p+\mu(\Tilde{p}a-\Tilde{a}a)\\
    &+\kappa(\Tilde{a}^2ap+\Tilde{a}ap-\Tilde{a}a\Tilde{p}p-a\Tilde{p}p)\nonumber
    \eea
At mean-field level, this gives the expected equations of motion, $\dot \rho_A = -\mu \rho_A+\kappa\rho_P\rho_A$ and $\dot \rho = 0$ \cite{SM}.  

{\em Gaussian Model:} Next we expand \eqref{eq:action} about the mean field solution and shift $a=a_0+\check{a}$ and $p=p_0+\check{p}$. Here $a_0$ and $p_0$ are the mean densities of a set of Poisson-distributed active and passive particles initialized in the distant past \cite{SM}. At Gaussian (linear) level, these densities remain unchanged under time evolution; consequently, 
$a_0 = \rho-\rho_{c,g}$ and $p_0 = \rho_{c,g} = \mu/\kappa$, where $\rho_{c,g}$ is the bare (Gaussian) value of the critical density. In particular, $a_0=0$ identifies the critical point. (Crucially, this feature of the Gaussian model will change below for the nonlinear theory.) The quadratic action density is thereby found as
$-\Tilde{a}(\partial_t-D\nabla^2)\check{a}-\Tilde{p}(\partial_t+\kappa a_0)\check{p}+\kappa a_0\Tilde{a}\check{p}+\kappa a_0p_0(\Tilde{a}^2-\Tilde{a}\Tilde{p})
$. Propagators can be read off from here, but in Doi-Peliti theory there is a nontrivial relation between terms in the action and the physical densities and noises \cite{Tauber,Biroli}. This means that calculating static density correlators $S_{\alpha\beta}(q)$ for $\alpha,\beta \in A,P$ requires a tree-level computation; see \cite{SM}. The results are
\beq
S_{AA} = a_0 +\frac{\kappa a_0 p_0}{Dq^2+\kappa a_0} \;;\;
S_{PP} = p_0 \;;\; S_{AP} = \frac{-\kappa a_0 p_0}{Dq^2+\kappa a_0}\nonumber\eeq
giving for the combined density-density correlator
\bea
\langle\rho_q\rho_{-q}\rangle &\equiv S(q)= S_{AA}+2S_{AP}+S_{PP}\label{eq:Stot}\\
&= a_0 + \frac{p_0Dq^2}{Dq^2+\kappa a_0} = a_0 + p_0\frac{q^2\xi^2}{1+q^2\xi^2} \label{eq:GSF}
\eea
The critical point is at $a_0\propto\xi^{-2}\to 0$. This implies vanishing of $S(0)$, and hence hyperuniformity, so long as the $q\to 0$ limit is taken first, whereas for finite $q$, $S(q)$ approaches a constant, $p_0$, as $a_0\to 0$. Therefore at criticality  $S(q)$ is zero at the origin but $p_0$ elsewhere, which can be formally viewed as $S(q) \sim q^\varsigma$ with exponent ${\varsigma = 0^+}$. (Note, however, that this order of limits reverses the one used in RG to access the critical scaling.) 

\begin{figure}
\includegraphics[width=0.365\columnwidth]{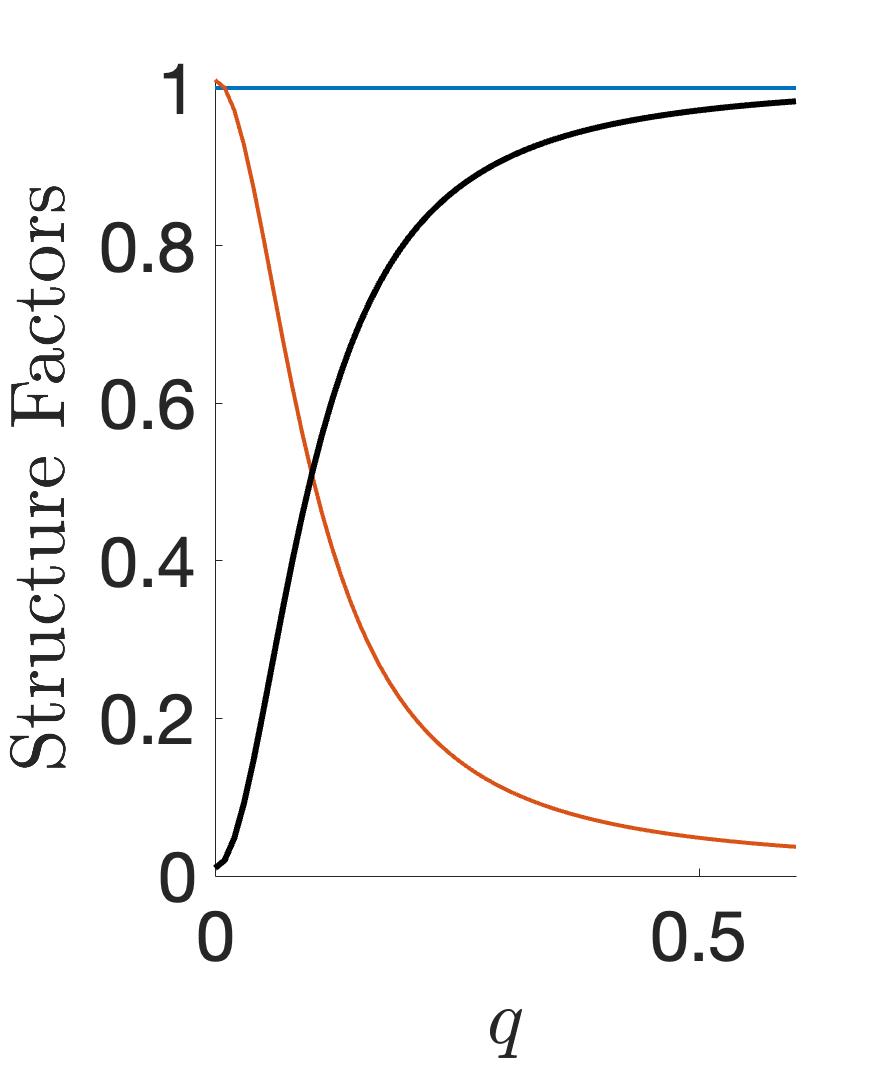}
\includegraphics[width=0.615\columnwidth]{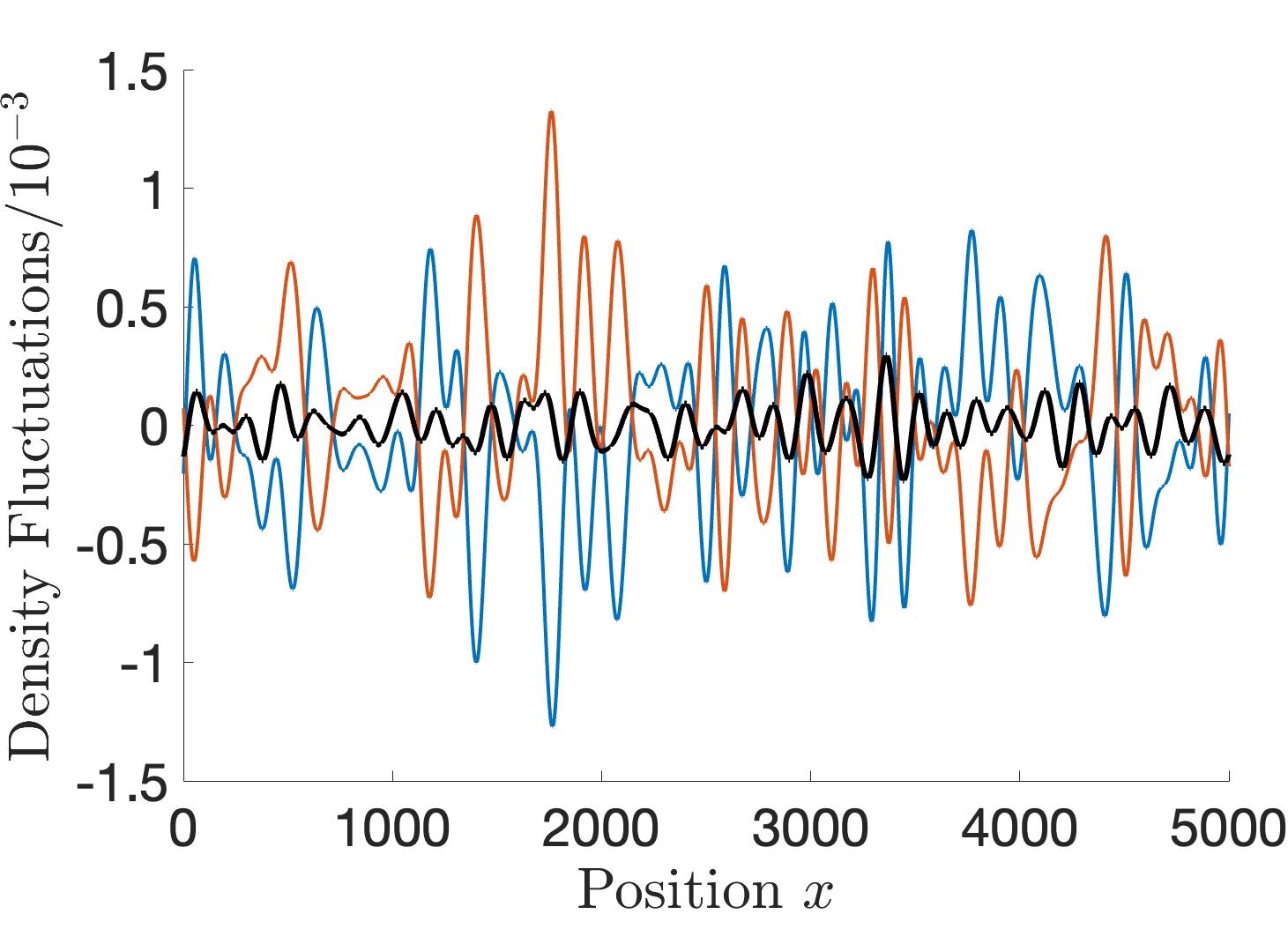}
\caption{(a) Plot of structure factors $S(q), S_{AA}(q), S_{PP}(q)$ vs $q$ for $a_0=0.01, p_0 =\kappa=D=1$ (giving $\xi=10$) showing at low $q$ the cancellation-induced suppression of  total density fluctuations. Blue line (horizontal) passive; red (decreasing) active; black (increasing) total density. (b) Sample of spatial density statistics for the Gaussian model in $d=1$. Parameters as for (a); bold black line is the total density.}
\label{fig:one}
\end{figure}

Fig.~\ref{fig:one} shows structure factors, and sampled density profiles (in $d=1$ for simplicity), at small $a_0$. 
While the Gaussian model does not enforce positivity of $\rho_A$ and $\rho_P$, it strikingly demonstrates how hyperuniformity emerges by anticorrelation of passive and active particles. This must be so, because the $S_{AA,PP}(0)$ correlators found above each remain finite at criticality ($a_0\to 0$), where the full density has $S(0)=0$. We show later that enforcing positivity of the cancelling densities requires strong non-Gaussianity only for $d<4$.

A further aspect of the Gaussian theory is exposed by using a Cole-Hopf transformation \cite{Biroli}  to find the equivalent Langevin equations, from which the same structure factors can alternatively be derived (see \cite{SM}): 
\begin{align}
    \frac{\partial \rho_A}{\partial t}&=D\nabla^2 \rho_A+\kappa a_0(\rho-\rho_A-p_0)\nonumber\\
    &+\sqrt{2\mu a_0}\eta+\sqrt{2Da_0}\nabla\cdot\boldsymbol{\Xi}\label{eq:CH1}\\
    \frac{\partial\rho}{\partial t}&=D\nabla^2\rho_A+\sqrt{2Da_0}\nabla\cdot\boldsymbol{\Xi}\label{eq:CH2}
\end{align}
Here $\eta$ and $\boldsymbol{\Xi}$ are unit white Gaussian noises, arising respectively from reactions and diffusion. From these equations one finds that without the diffusive noise $\boldsymbol{\Xi}$, the $a_0$ term in \eqref{eq:GSF} is absent: hyperuniformity (now with $\varsigma = 2$) then also arises at $\xi<\infty$, that is, {\em away from} criticality. 

Thus, neglecting diffusive noise alters the Gaussian-level predictions dramatically, although it is often considered subdominant for C-DP \cite{Janssen,Wiese,Sriram}. The exponent $\varsigma=2$ throughout the active phase matches results for particles with C-DP-like interactions that conserve centre of mass \cite{HexnerCoM}. Notably, without noise, \eqref{eq:CH2} has an equivalent conservation law, as follows. The `mass-moment' density field ${\bf p} = \rho{\bf r}$ obeys $\dot p_\alpha = -(\nabla_\beta J_\beta)r_\alpha = -\nabla_\beta (J_\beta r_\alpha)+ J_\alpha$. For $\boldsymbol{\Xi}=0$ only, the current  ${\bf J}= -D\nabla\rho_A$ in \eqref{eq:CH2} is of pure gradient form, so that $\dot p_\alpha = - \nabla_\beta\Sigma_{\beta\alpha}$ with ${\bf \Sigma}={\bf J r}+{\bf I}D\rho_A$ where ${\bf I}$ is the unit tensor. Hence the mass-moment $\int{\bf p}d{\bf r}$ in a local region is conserved unless there are currents ${\bf \Sigma}$ across its boundary \cite{Tomita}. We focus below on cases {\em without} this additional conservation law.

{\em RG Analysis:} The Gaussian Model breaks down below the upper critical dimension which, after earlier doubt \cite{Fred}, was found via the q-EW mapping as $d_c = 4$ \cite{Doussal}. Our RG analysis starts from the action \eqref{eq:action} working to one loop, and finds exponents to order $\epsilon$.  This follows broadly standard lines \cite{Tauber,DP}, with two nonstandard features, summarized below and detailed in \cite{SM}. 

\begin{figure}
\includegraphics[width=0.8\columnwidth]{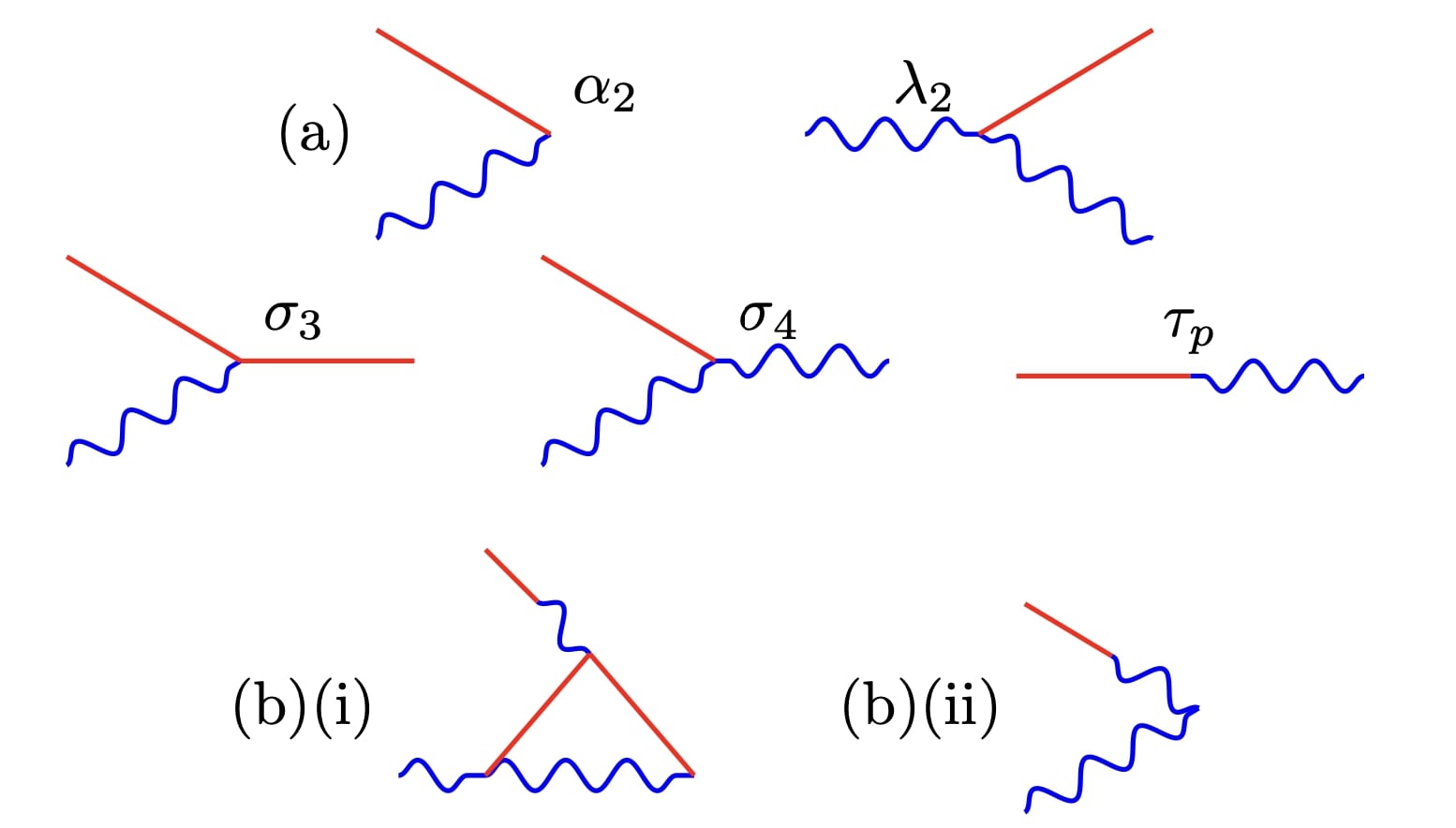}
\caption{(a) Vertices whose renormalization defines the three independent coupling constants of the theory (see text); their symmetric counterparts $\alpha_1$, $\sigma_{1,2}$, $\lambda_1$ replace the left wavy lines with a straight one.  Straight lines are active fields, wavy lines passive ones, and time proceeds leftwards so that, {\em e.g.}, $\tau_p$ represents conversion of a passive to an active particle at a rate $\tau_p$ (whose bare value is $\kappa a_0$). (b)(i) An example of additional loop diagram contributing at $\mathcal{O}(\epsilon)$ despite the transmutation arising outside the loop \cite{GunnarPC}, since its equivalent tree-level correction, shown in (b)(ii), cannot be generated by usual tree-level expansion. For details see \cite{SM}.}
\label{fig:two}
\end{figure}

The first nonstandard feature of the perturbative field theory is that classification of 1-particle reducible and irreducible diagrams requires particular care \cite{SM}. In multi-species reaction-diffusion systems, transmutation vertices such as $\tau_p$ in Fig.~\ref{fig:two}(a) are often present. Naively interpreting them as interaction vertices rules out loop corrections with transmutations outside the loop, and therefore misses out corrections to vertices that have zero bare value and that cannot be generated by usual tree-level expansions \cite{GunnarPC}; see  Fig.~\ref{fig:two}(b). Moreover, by viewing transmutations as part of the propagators in the bilinear theory when necessary, we can identify several symmetries in pairs of interaction vertices, thereby reducing the number of independent coupling constants. Using particle conservation, these symmetries can be derived independently and non-perturbatively \cite{SM}.

Carefully implementing such protocols, we find that certain combinations of vertices, equal and opposite at bare level, remain so under RG flow. Three independent coupling constants remain, denoted $u,v,w$, controlling respectively the combinations $\lambda_1\sigma_3$, $\lambda_2\sigma_4$ and $\lambda_2^2\alpha_2\tau_p/\epsilon_p^2$ of the six vertices shown in Fig.~\ref{fig:two}(a). 
(Here $\epsilon_p$ is the mass term in the passive propagator, controlling the distance from criticality.) 
Constructing the RG flow equations to one loop, the fixed-point values of these couplings are found as $u^* = v^* = -2\epsilon/9$ and $w^* = 2\epsilon/3$ \cite{SM}. This is enough to determine two critical exponents, via $z=2+u^*$ and $1/\nu_\perp = 2-w^*/2$, whose order-$\epsilon$ values agree with those from the q-EW mapping \cite{Doussal}.

The second nonstandard feature of our RG calculations is an ambiguity in the scaling dimensions of the fields $\check{a},\Tilde a, \check{p}, \Tilde p$. (This must be resolved in order to find the remaining exponents $\beta$ and $\varsigma$.)
The ambiguity shows up even at the level of bare dimensions, as we now describe. Under length rescaling $x\to x/\zeta$ we define $[u] = y$ for variable $u$ if $u\to u\zeta^y$; thus $[x]=-1, [q] = 1$. From \eqref{eq:action} it then follows that $[\Tilde p \check p] = [\Tilde p] +[\check p] = d$; likewise for $a$. Density conservation also requires $[\check a]=[\check p]$. In some models with a single absorbing state, a `rapidity reversal' symmetry \cite{Lubeck} in a certain species such as the passive fields, $\Tilde p(x,t)\leftrightarrow -\check p(x,-t)$, requires  $[\check p] = [\Tilde p] = d/2$; together these considerations would also imply $[\check a] = [\Tilde a] = d/2$.  However, for C-DP no rapidity-reversal symmetry is apparent in \eqref{eq:action}, creating a technical impasse. 
This does not affect $\nu_\perp$ and $z$ above, because the calculations used to find those involved only the field products $\tilde a \check a$ and $\tilde p \check p$. Under RG, these acquire anomalous dimensions such that $[\Tilde a\check a] = d+ 2\epsilon/9$ and $[\Tilde p\check p] = d-\epsilon/9$ \cite{SM}. 

To find the remaining exponents $\beta$ and $\varsigma$ we need the dimensions of each field separately.   Below we use results given in \cite{SM} to show that only one choice (the same as would arise from rapidity reversal in passive particles) is consistent with hyperuniformity, with all other choices giving {\em divergent} rather than zero  low-$q$ fluctuations at criticality.  It is, of course, not unusual for physical knowledge concerning a critical point to resolve ambiguities in an RG calculation, but intriguing that here the requirement of hyperuniformity itself is sufficient to do so.

To see how this works, let us consider $S(q)$. In direct counterpart to \eqref{eq:GSF} we find at one loop level, after removal of terms that vanish or cancel without any assumption concerning the field dimensions, an expression that mirrors the form of \eqref{eq:Stot} for the Gaussian case \cite{SM}:
\begin{equation}
S(q) = a_0\langle \check a\Tilde a\rangle + \langle \check a \check a\rangle +
\langle \check a \check p\rangle + \langle \check p \check a\rangle + \langle \check p \check p\rangle+ p_0\langle \check p \Tilde p\rangle
\label{eq:fieldS}
\end{equation}
Here all the correlators are equal-time, with $q$ arguments suppressed for clarity. The first two terms represent the active-active density correlator, the next two the active-passive cross correlations, and the final term the passive-passive density correlator.  Our RG approach gives no information on amplitude ratios for these terms, so that any cancellations among them cannot be directly established at that level (unlike the Gaussian case). However, their $q$ dependences at criticality are directly set by the scaling dimensions of the four fields, $\check a,\Tilde a, \check p, \Tilde p$, and this will be enough for us.

From the anomalous dimensions reported above for $[\check a\Tilde a],[\check p\Tilde p]$, we observe that at criticality ($\xi\to\infty$) the first of the six terms in \eqref{eq:fieldS} scales  as $S_1 \sim a_0q^{2\epsilon/9}$ and the last as $S_6\sim p_0q^{-\epsilon/9}$.  Note that, as mentioned previously, the shift $a_0$ no longer vanishes at criticality as it does in the Gaussian limit. Moreover $a_0$ cannot be written as a combination of the $u,v,w$ effective coupling constants, nor of the particle field operators, arising in the action \eqref{eq:action} or its shifted counterpart \cite{SM}. This means that $a_0$ cannot acquire an anomalous dimension under the RG flow: it merely acts as a non-universal amplitude. (The same is true of $p_0$, as can separately be confirmed by requiring $\beta$ to match the q-EW result.) 
Accordingly, for the system to be hyperuniform rather than divergently fluctuating at low $q$, $S_6$ must be cancelled by some combination of the terms $S_{2,3,4,5}$. This requirement {\em alone} fixes the anomalous dimensions of the fields as $\eta_{\check a} = \eta_{\check p}=\eta_{\Tilde p} = -\epsilon/18$ and $\eta_{\Tilde a} = 5\epsilon/18$: only then   can all negative powers of $q$ cancel in \eqref{eq:fieldS} \cite{SM}.

Although the terms involved are now divergent rather than finite at $q\to 0$, this cancellation resembles the one found (albeit via a different order of limits) for the Gaussian model 
in (\ref{eq:Stot},\ref{eq:GSF}) and Fig.~\ref{fig:one}. As found there, strongly fluctuating active and passive quantities must cross-correlate such that their sum is hyperuniform.
Moreover, since every term $S_{i>1}$ in \eqref{eq:fieldS} involve just two fields, each has a pure scaling behavior $\xi^0q^{-\epsilon/9}F_i(q\xi)$ with $F_i(s)$ regular at large $s$ \cite{Tauber}. Hyperuniformity then requires $\sum_{i=2}^6 F_i(\infty) = 0$, and whatever remains after this cancellation vanishes at criticality where $\xi\to\infty$. (This reasoning would not hold if the individual $S_i$ were, like $S$, correlators of sums of fields \cite{Amit}.) Accordingly, given that $a_0$ is not singular as detailed above, the hyperuniformity exponent governing $S(q)$ at the critical point  can be read off from $S_1$ as $\varsigma = 2\epsilon/9$.

As we have emphasised, the critical regime exhibits hyperuniformity of the total density $\rho$, but not of the active and passive densities separately. Instead, the correlators for these each diverge as $q^{-\epsilon/9}$, confirming a previously known value of $2-\epsilon/9$ \cite{Doussal} for the exponent $\eta_\perp$ defined  via $S_{AA} \sim = q^{-2+\eta_\perp}$ \cite{Lubeck}. The anomalous dimensions determined above also imply $\beta = 1-\epsilon/9$ \cite{SM}. Indeed, an alternative way of fixing those dimensions is to impose this value of $\beta$, already known from the q-EW mapping \cite{Doussal}; hyperuniformity with $\varsigma = 2\epsilon/9$ then follows. (A further route to this answer would be to ansatz an emergent rapidity reversal symmetry.)

{\em Discussion:} 
The near perfect cancellation of active and passive density fluctuations on approach to criticality is remarkable since the mean density of active particles itself vanishes at the critical point. One may ask: how can fluctuations among a vanishingly small density of active particles perfectly cancel those of a nonvanishing density of passive particles whose fluctuations are either finite (in $d>4$, where $S_{PP} = p_0$) or even divergent (in $d<4$, where $S_{PP}\sim q^{-\epsilon/9}$)?

The Gaussian case is again instructive. Here, the ideal-gas-like structure factor for passive particles, $S_{PP} = p_0$, implies that in a cube of side $\lambda$  such that the passive particle number has mean $N_P(\lambda) = p_0\lambda^d \gg 1$, its standard deviation obeys $\sigma_P(\lambda)^2\sim p_0\lambda^d$. To cancel the (Gaussian) fluctuations in passive density requires active particles to have the same standard deviation $\sigma_A(\lambda)  = \sigma_P(\lambda)$, but now with a mean of only $N_A(\lambda)= a_0\lambda^d$. This can be done, with near-Gaussian fluctuations and without creating negative $\rho_A$ locally, only if $\sigma_A(\lambda) \lesssim N_A(\lambda)$. This requires $p_0^{1/2}\lambda^{d/2}  \lesssim a_0\lambda^d $, where $a_0^{-1}\sim \xi^2$, and hence
$\lambda \gtrsim \xi^{4/d}$. (Both lengths are here measured in microscopic units.) Hence near-Gaussian fluctuations of the minority active particles {\em can} cancel the majority passive fluctuations at scales $\lambda \gtrsim \xi$ (which is where hyperuniformity sets in at Gaussian level, see \eqref{eq:GSF}), but only if $d>4$. In lower dimensions this is not possible. This gives new  insight into C-DP's upper critical dimension, $d_c = 4$. 

In $d<4$,  the mean number of active particles in a box of size $\lambda$ now varies as $N_A(\lambda) \sim \lambda^d\rho_A \sim \lambda^d \xi^{-\beta/\nu_\perp}\sim \lambda^{4-\epsilon}\xi^{-2+5\epsilon/9}$, whereas the variance must obey $\sigma_A(\lambda)^2=\sigma_P(\lambda)^2 \sim S_{PP}(\lambda^{-1})\lambda^d$, giving $\sigma_A(\lambda) \sim \lambda^{2-4\epsilon/9}$. (This estimate follows from the usual relation between compressibility and structure factor, now applied to a subsystem of size $q^{-1}=\lambda$.) Requiring $N_A(\lambda)\gtrsim \sigma_A(\lambda)$ as before yields $\lambda^{2-5\epsilon/9}\gtrsim\xi^{2-5\epsilon/9}$ and hence $\lambda\gtrsim\xi$. 
This marginal outcome can be extended beyond order $\epsilon$ by use of the scaling relation $2-\eta_\perp -d = -\beta/\nu_\perp$ \cite{Lubeck}. It confirms that in $d<4$ significantly non-Gaussian fluctuations of $\rho_A$ are needed to avoid negative values: the standard deviation in active particle number is of order its mean in a correlation-length sized box. Without proving that the argument extends across a cascade of shorter scales, as our RG results say it must, we think this makes `hyperuniformity by cancellation' less mysterious.

A second striking feature of our central result, $\varsigma = 2\epsilon/9$, is that it disproves a previous analytical prediction for the hyperuniformity exponent of the C-DP class. This prediction was conjectured by Hexner and Levine \cite{HexLev} via a proposed scaling law
\beq
\varsigma = 2-\eta_\perp = \epsilon/9+\mathcal{O}(\epsilon^2)\label{eq:HL}
\eeq
(Their notation has $\varsigma \equiv \lambda-d$.)
It is not without precedent for a closed-form prediction for a critical exponent to fail close to $d=4$, and thus be disproved in general, despite being usefully accurate in lower dimensions \cite{Foot1}. In fact though, our predictions using $\varsigma=2\epsilon/9$ (namely $\varsigma=(0.22,0.44,0.66)$ in $d = (3,2,1)$) are almost as close to the numerical results of \cite{HexLev},  $\varsigma  = (0.24,0.45, 0.43)$) as are the predictions from \eqref{eq:HL}, even if the $\eta_\perp(d)$ values derived from numerical observations are used there. (The latter procedure gives $\varsigma = (0.26,0.46,0.58)$.) Certainly our $\varsigma = 2\epsilon/9$ lies closer to the data than the $\mathcal{O}(\epsilon)$ expansion of \eqref{eq:HL}, which yields $\varsigma=(0.11,0.22,0.33)$. 

A further notable feature of our work concerns the C-DP exponents ($\beta, \nu_\perp$ and $z$) previously found via the mapping to q-EW, whose treatment to order $\epsilon$ required FRG methods  \cite{qEWexponents}. On this basis one might have guessed that FRG was an essential tool to address the C-DP class; yet we have calculated all its exponents without FRG here. Clearly, for two physically distinct models within the same universality class (reaction-diffusion and interfacial depinning), FRG may be needed for one and not the other. More generally, studying more than one model in a class may allow the full set of exponents to be found sooner. Indeed for C-DP, we are not sure whether the exponent $\varsigma$ could ever be calculated in the q-EW setting. This could be especially challenging if the subleading (conserved) noise term is essential to avoid an unwanted conservation law, as we found it to be at Gaussian level. 

In summary, the hyperuniformity exponent $\varsigma$ describes a signature elements of the C-DP universality class, manifested in the physics of random organization for dilute colloids  \cite{RO1,RO2,RO3}, in similar transitions at high density in colloids and granular media \cite{Sood,Royer,Ness}, and in other reaction-diffusion processes with many absorbing states. Our calculation of this exponent to order $\epsilon = 4-d$ has shed light on many aspects of C-DP physics, including: the role of conservative noise; a form of hyperuniformity in $d>4$; and the way hyperuniformity emerges via near-perfect anticorrelation of active and passive densities that are not separately hyperuniform but have finite (for $d>4$) or divergent (for $d<4$) fluctuations. We believe these to be significant advances towards a more complete understanding of C-DP physics, and hope they will drive further numerical and experimental investigations of this important class of problems.

{\em Acknowledgments:} We thank Marius Bothe, Cathelijne ter Burg, Cesare Nardini, Gunnar Pruessner and Frederic van Wijland for helpful discussions. We also thank Marius Bothe and Gunnar Pruessner specifically for valuable insights concerning the treatment of zero bare vertices the loop expansion \cite{GunnarPC}. Work funded in part by the European Research Council under the Horizon 2020 Programme, ERC Grant Agreement No. 740269. XM thanks the Cambridge Commonwealth, European and International Trust and China Scholarship Council for a joint studentship. JP was supported through a UKRI Future Leaders Fellowship (MR/T018429/1 to Philipp Thomas).


\end{document}


\title{Supplemental Material: Theory of Hyperuniformity at the Absorbing State Transition}
\author{Xiao Ma}
\affiliation{DAMTP, Centre for Mathematical Sciences, University of Cambridge, Wilberforce Road, Cambridge, CB3 0WA}
\author{Johannes Pausch}
\affiliation{DAMTP, Centre for Mathematical Sciences, University of Cambridge, Wilberforce Road, Cambridge, CB3 0WA}
\affiliation{Department of Mathematics, Imperial College, London SW7 2AZ}
\author{Michael E. Cates}
\affiliation{DAMTP, Centre for Mathematical Sciences, University of Cambridge, Wilberforce Road, Cambridge, CB3 0WA}

\date{\today}
\maketitle
\onecolumngrid
In the supplemental material we show full theoretical calculations we used to obtain results presented in the paper. In particular, we address several subtleties in greater detail. In Section~\ref{sec:Derivation} the derivation of the Doi-Peliti action is outlined. The physical interpretation of a shift in annihilation fields in the Doi-Peliti formalism is explained in Section \ref{subsec:Initializing}. Section~\ref{sec:Mean} presents the mean-field theory based on our action.  In Section \ref{sec:Gaussian} we calculate structure factors in the Gaussian limit and  explain the application of the Cole-Hopf transformation in deriving diffusional noise. Section~\ref{sec:RG} outlines the renormalization group analysis, including discussion of the two non-standard features in our calculations mentioned in the main text. In Section~\ref{sec:Cancellations}, we explain how cancellations in the structure factor settle the anomalous dimensions of active and passive fields and, finally, the hyperuniformity exponent for C-DP.
\tableofcontents

\section{Derivation of the Doi-Peliti Action}\label{sec:Derivation}
In this Section we briefly show how the Doi-Peliti action can be derived from a master equation which describes the reaction processes  $A+P\to 2A$ with rate $\kappa$, $A\to P$ with rate $\mu$, and diffusion for active particles. Passive particles remain stationary. Letting $n_i$ denote the number of active particles at position $i$ and $m_j$ the number of passive particles at position $j$,  the master equation is
\begin{equation}
\begin{split}
\partial_t P(n,m,t)=&\,\mu\sum\limits_i\Bigl((n_i+1)P(n+1_i,m-1_i,t)-n_iP(n,m,t)\Bigr)\\
&\,+\kappa\sum\limits_i\Bigl((n_i-1)(m_i+1)P(n-1_i,m+1_i,t)-n_im_iP(n,m,t)\Bigr)\\
&\,+\text{diffusion}
\end{split}
\end{equation}
Here $n$ and $m$ are shorthands for the collections of all $n_i$ and $m_j$ respectively, and where $1_i$ is used to add or subtract a single particle at position $i$. Here and below, we do not write out the standard term for the diffusive hopping of (active) particles \cite{Tauber}.

Next, we introduce states $|n,m\rangle$ as well as ladder operators $a_i^\dagger$ and $a_i$ for active particles and $p_j^\dagger$ and $p_j$ for  passive particles with commutation relation $c_id_j^\dagger-c_i^\dagger d_j=\delta_{ij}\delta_{cd}$, ($c,d\in\{a,p\}$). The ladder operators act on the states by increasing particle numbers, $a_i^\dagger|n,m\rangle=|n+1_i,m\rangle$, or decreasing them, $a_i|n,m\rangle=n_i|n-1_i,m\rangle$. Now, the master equation above can be second-quantized into a Schroedinger-like equation for the system state $|\mathcal{M}(t)\rangle=\sum_{n,m}P(n,m,t)|n,m\rangle$:
\begin{align}
\partial_t|\mathcal{M}(t)\rangle=&\,\sum\limits_i\Bigl(\mu(p^\dagger_ia_i-a^\dagger_ia_i)+\kappa(a_i^{\dagger 2}a_ip_i-a^\dagger_ia_ip^\dagger_ip_i)+\text{diffusion}\Bigr)|\mathcal{M}(t)\rangle\\
=&\,\sum\limits_i\Bigl(\mu(\widetilde p_i-\widetilde a_i)a_i+\kappa(\widetilde a_i^2a_ip_i+\widetilde a_ia_ip_i-\widetilde a_ia_i\widetilde p_ip_i-a_i\widetilde p_ip_i)+\text{diffusion}\Bigr)|\mathcal{M}(t)\rangle
\end{align}
where in the last line, we have applied the `Doi-shifts' $a^{\dagger}=\Tilde{a}+1$, $
p^{\dagger}=\Tilde{p}+1$.

Using the coherent-state path integral representation \cite{Tauber}, we now write down the corresponding Doi-Peliti action density (Eq.1 of the main text)
\begin{equation}\label{eq:action}
\mathbb{A}=-\widetilde a(\partial_t-D\Delta+\mu)a-\widetilde p\partial_t p+\mu\widetilde pa+\kappa(\widetilde a^2ap+\widetilde aap-\widetilde aa\widetilde pp-a\widetilde pp)
\end{equation}
with diffusion constant $D$. All operators are now defined in continuous space (and time) rather than on a lattice.

One common point of confusion in Doi-Peliti field theory is that the annihilation fields $a$ and $p$ are not equal to the active and passive particle density fields $\rho_A$ and $\rho_P$. Although their first moments agree, $\langle a\rangle=\langle\rho_A\rangle$ and $\langle p\rangle=\langle\rho_P\rangle$, their higher moments do not. For instance, $\langle\rho_A(x,t)\rho_A(y,t)\rangle=\langle a(x,t)a(y,t)\rangle+\langle a(x,t)\Tilde{a}(y,t)a(y,t)\rangle$. Calculating the structure factor hence requires evaluating several observables in terms of the Doi-Peliti fields. 

Finally, we follow the established convention representing terms in the action and in the calculations of observables by Feynman diagrams. In Doi-Peliti field theory, Feynman diagrams are drawn according to the following rules:
\begin{itemize}
    \item Each annihilation field $a$ (or $p$) in the calculation of observable $\mathcal{O}$ is represented as a left end point of a leg.
    \item Each creation field $\Tilde{a}$ (or $\Tilde{p}$) in the calculation of observable $\mathcal{O}$ is represented as a right end point of a leg.
    \item Each interaction vertex in the action $\Tilde{a}^ka^l\Tilde{p}^mp^n$ is represented by a node with $l$ active propagators and $n$ passive propagators coming from the right, and $k$ active propagators and $m$ passive propagators going out from the left.
    \item Time flows from right to left, and hence the left-right order of lines must be obeyed in order to respect time ordering.
\end{itemize}
We draw active fields as straight red lines and passive fields as blue wavy lines.

\section{Initializing the system -- shifting the annihilation fields}\label{subsec:Initializing}

In order to study hyperuniformity, we need to initialize the system with active and passive densities close to the critical point. In Doi-Peliti field theory, this can be efficiently achieved by shifting the annihilation fields, as will be explained in the following. 

Focusing only on active particles for now, an $a_0$-Poisson distribution at site $i$ can be initialized at time $t_0$ with the help of creation operators:\begin{equation}
    |\mathcal{M}(t_0)\rangle=e^{-a_0}\sum_{l=0}^{\infty}\frac{a_0^l}{l!}|l_i,0\rangle=e^{-a_0}\sum_{l=0}^{\infty}\frac{a_0^la_i^{\dagger l}}{l!}|0,0\rangle
\end{equation}
This initialization can be applied at every position and in order to shorten the notation, we drop the subscript $i$.

If the expectation of an observable $\mathcal{O}$ is evaluated at time $t_1>t_0$, the initialization can be compactly written as an additional term that is added to the action:\begin{align}
    \left\langle\mathcal{O}(t_1)e^{-a_0}\sum_{l=0}^{\infty}\frac{a_0^la^{\dagger l}(t_0)}{l!}\right\rangle&=\int\mathcal{D}[\Tilde{a},a]\mathcal{O}(t_1)e^{\mathcal{A}+a_0(a^{\dagger}(t_0)-1)}\\
    &=\int\mathcal{D}[\Tilde{a},a]\mathcal{O}(t_1)e^{\mathcal{A}+a_0\Tilde{a}(t_0)}
\end{align}
This additional term can be absorbed in the action by shifting the annihilation field by $a_0$ from time $t_0$ onwards, $a=\check a+a_0\Theta(t-t_0)$: $-\Tilde{a}\partial_t a$ is replaced by $-\Tilde{a}\partial_t\check a-a_0\Tilde{a}\delta(t-t_0)$.
After applying integration of the action density in time, the second term cancels with the extra $a_0\Tilde{a}(t_0)$ created by the initialization.

Finally, we push back the initialization time $t_0\rightarrow-\infty$ which makes the Heaviside function $\Theta(t-t_0)$ obsolete. The same procedure can be used for initializing the passive particles with a spatially uniform $p_0$-Poisson distribution.

We use the annihilation field shifts in our theory and choose $p_0=\mu/\kappa$, eliminating $\mu$ from the action density:
\begin{multline}\label{eq:shiftedAction}
    \mathbb{A}= -\Tilde{a}(\partial_t-D\nabla^2)\check a-\Tilde{p}(\partial_t+\kappa a_0)\check p+\kappa a_0\Tilde{a}\check p+\kappa a_0p_0(\Tilde{a}^2-\Tilde{a}\Tilde{p})\\
    +\kappa p_0(\Tilde{a}^2-\Tilde{a}\Tilde{p})\check a+\kappa a_0(\Tilde{a}^2-\Tilde{a}\Tilde{p})\check p+\kappa(\Tilde{a}-\Tilde{p})\check a\check p+\kappa(\Tilde{a}^2-\Tilde{a}\Tilde{p})\check a\check p 
    \end{multline}
We introduce new names for all coupling constants in order to organize them systematically according to topology. The passive-to-active transmutation coupling will be denoted by $\tau_p$ and the passive mass by $\epsilon_p$. The active-to-passive transmutation coupling $\tau_a$ has zero bare value as does the active bare mass $\epsilon_a$. We represent interaction vertices as amputated Feynman diagrams (shown with their bare values):
\begin{align}\label{eq:vertexDiagrams}
\begin{aligned}
\tikz[thick]{\draw[color=red] (-1,-0.6)--(0,0) node[above right,black] {$\alpha_1=\kappa p_0a_0$};\draw[color=red] (-1,0.6)--(0,0) ;}\qquad\tikz[thick]{\draw[color=red] (-1,-0.6)--(0,0) node[above right,black] {$\sigma_1=\kappa p_0$};\draw[color=red] (-1,0.6)--(0,0);\draw[color=red] (0,0) -- (0.6,0);}\qquad\tikz[thick]{\draw[color=red] (-1,-0.6)--(0,0) node[above right,black] {$\sigma_2=\kappa a_0$};\draw[color=red] (-1,0.6)--(0,0);\draw[color=blue,decorate,decoration=snake] (0,0) -- (0.8,0);}\qquad\tikz[thick]{\draw[color=red] (-1,0.0)--(0,0) node[above left,black] {$\lambda_1=\kappa$};\draw[color=blue,decorate,decoration=snake] (1,-0.6) -- (0,0);\draw[color=red] (0,0) -- (1,0.6);}\qquad\tikz[thick]{\draw[color=red] (-1,-0.6)--(0,0) node[right=0.3cm,black] {$\chi_1=\kappa$};\draw[color=red] (-1,0.6)--(0,0);\draw[color=blue,decorate,decoration=snake] (1,-0.6) -- (0,0);\draw[color=red] (0,0) -- (1,0.6);}\\
\tikz[thick]{\draw[color=blue,decorate,decoration=snake] (-1,-0.6)--(0,0) node[above right,black] {$\alpha_2=-\kappa a_0p_0$};\draw[color=red] (-1,0.6)--(0,0);}\qquad
\tikz[thick]{\draw[color=blue,decorate,decoration=snake] (-1,-0.6)--(0,0) node[above right,black] {$\sigma_3=-\kappa p_0$};\draw[color=red] (-1,0.6)--(0,0);\draw[color=red] (0.6,0.0) -- (0,0);}\qquad\tikz[thick]{\draw[color=blue,decorate,decoration=snake] (-1,-0.6)--(0,0) node[above right,black] {$\sigma_4=-\kappa a_0$};\draw[color=red] (-1,0.6)--(0,0);\draw[color=blue,decorate,decoration=snake] (0.8,0.0) -- (0,0);}\qquad\tikz[thick]{\draw[color=blue,decorate,decoration=snake] (-1,0.0)--(0,0) node[above left,black] {$\lambda_2=-\kappa$};\draw[color=blue,decorate,decoration=snake] (1,-0.6) -- (0,0);\draw[color=red] (0,0) -- (1,0.6);}\qquad
\tikz[thick]{\draw[color=blue,decorate,decoration=snake] (-1,-0.6)--(0,0) node[right=0.3cm,black] {$\chi_2=-\kappa$};\draw[color=red] (-1,0.6)--(0,0);\draw[color=blue,decorate,decoration=snake] (1,-0.6) -- (0,0);\draw[color=red] (0,0) -- (1,0.6);}
\end{aligned}
\end{align} 
In Subsection~\ref{subsec:1-PI} on renormalization, other interaction vertices with coupling constants of zero bare value are important in establishing symmetries. Although they don't appear in the bare action~\eqref{eq:shiftedAction}, we list them here to complete the overview of interaction vertices and names for couplings constants:\begin{align}
\tikz[thick]{\draw[color=blue,decorate,decoration=snake] (-1,-0.6)--(0,0) node[above right,black] {$\alpha_3=0$};\draw[color=blue,decorate,decoration=snake] (-1,0.6)--(0,0);}\qquad\tikz[thick]{\draw[color=blue,decorate,decoration=snake] (-1,-0.6)--(0,0) node[above right,black] {$\sigma_5=0$};\draw[color=blue,decorate,decoration=snake] (-1,0.6)--(0,0);\draw[color=red] (0.6,0.0) -- (0,0);}\qquad\tikz[thick]{\draw[color=blue,decorate,decoration=snake] (-1,-0.6)--(0,0) node[above right,black] {$\sigma_6=0$};\draw[color=blue,decorate,decoration=snake] (-1,0.6)--(0,0);\draw[color=blue,decorate,decoration=snake] (0.8,0.0) -- (0,0);}
\end{align}
There are also another 4 vertices of the $\lambda$ type:\begin{align}
\tikz[thick]{\draw[color=red] (-1,0.0)--(0,0) node[above left,black] {$\lambda_3=0$};\draw[color=red] (1,-0.6) -- (0,0);\draw[color=red] (0,0) -- (1,0.6);}\qquad
\tikz[thick]{\draw[color=blue,decorate,decoration=snake] (-1,0.0)--(0,0) node[above left,black] {$\lambda_4=0$};\draw[color=red] (1,-0.6) -- (0,0);\draw[color=red] (0,0) -- (1,0.6);}\qquad
\tikz[thick]{\draw[color=red] (-1,0.0)--(0,0) node[above left,black] {$\lambda_5=0$};\draw[color=blue,decorate,decoration=snake] (1,-0.6) -- (0,0);\draw[color=blue,decorate,decoration=snake] (0,0) -- (1,0.6);}\qquad
\tikz[thick]{\draw[color=blue,decorate,decoration=snake] (-1,0.0)--(0,0) node[above left,black] {$\lambda_6=0$};\draw[color=blue,decorate,decoration=snake] (1,-0.6) -- (0,0);\draw[color=blue,decorate,decoration=snake] (0,0) -- (1,0.6);}
\end{align}
The $\lambda_5$ and $\lambda_6$ are generated at zero-loop level via transmutations. In contrast $\lambda_3$ and $\lambda_4$ are generated only at one loop level.
Furthermore, our choice of $p_0=\mu/\kappa$ implies that the two source terms have zero bare value, and likewise emerge at one loop:
\begin{align}
\tikz[thick]{\draw[color=red] (-1,0.0)--(0,0) node{$\times$} -- (0,0) node[right,black] {$\gamma_1=0$};}\qquad\tikz[thick]{\draw[color=blue,decorate,decoration=snake] (-1,0.0)--(0,0) node{$\times$} -- (0,0) node[right,black] {$\gamma_2=0$};}
\end{align}
Finally, there are also another 7 interaction vertices of the $\chi$ type. However, all $\chi$ type vertices turn out to be irrelevant, in the usual RG sense, in dimensions $d>2$; see Subsection~\ref{subsec:Rescaling}.

\section{Mean-Field Approximation}\label{sec:Mean}
We start with the mean-field approximation, where we assume that active and passive particle densities, $\rho_A$ and $\rho_P=\rho-\rho_A$, are spatially uniform and fluctuations are absent. The equations of motion derived in this approximation are deterministic and can be found by minimising the action \eqref{eq:action} with respect to the Doi-shifted creation fields. We identify the active particle annihilation field that minimises the action with the mean-field density $\rho_A$ and the minimising passive particle annihilation field with $\rho_P$. This gives
\begin{align}\label{eq:meanfield}
    \frac{\partial \rho_A}{\partial t}&=-\mu \rho_A+\kappa (\rho-\rho_A)\rho_A\\
    \frac{\partial \rho}{\partial t}&=0
\end{align}

The total density of active and passive particles $\rho(x,t)=\rho_A(x,t)+\rho_P(x,t)$ does not evolve and acts as a control parameter of the system. When $\rho$ is larger than a critical value $\rho_c=\mu/\kappa$, there are two stationary solutions to the mean-field equations, an unstable one where $\rho_A=0$ (the absorbing state), and the other stable solution with $\rho_A=(\rho-\rho_c)^1>0$, $\rho_P=\rho_c$. At mean-field level, this says that the system stays in the active phase indefinitely. As the total density is decreased below the critical value $\rho_c$, the only one stable solution is $\rho_A=0$, and the system is in the absorbing phase. Thereby the well-known mean-field critical exponent $\beta = 1$ \cite{Lubeck} is confirmed.

\section{Gaussian Approximation}\label{sec:Gaussian}
For the Gaussian approximation, active and passive particle densities are assumed to be close to their mean-field values, ignoring all terms quadratic or higher in their perturbations. In contrast to the mean-field approximation above, active particles diffuse in the Gaussian approximation. To implement this field-theoretically, we need two steps: one, finding the Gaussian part of the action, and two, translating physical observables into diagrams constructed out of the Gaussian action.

In the Gaussian approximation, only terms at or below quadratic order are considered, i.e.~only the first column of interaction vertices in Eq.~\eqref{eq:vertexDiagrams} above, called noise vertices, remain in the action: 
\begin{equation}\label{eq:GaussianAction}
    \mathbb{A}_{\text{G}}=-\Tilde{a}(\partial_t-D\nabla^2)\check a-\Tilde{p}(\partial_t+\kappa a_0)\check p+\kappa a_0\Tilde{a}\check p+\kappa a_0p_0(\Tilde{a}^2-\Tilde{a}\Tilde{p})
    \end{equation}
Using this action, we observe that the {\em mean} densities $a_0+\langle\check a\rangle=\langle\rho_A\rangle$ and $p_0+\langle \check p\rangle=\langle\rho_P\rangle$ remain conserved under time evolution in the Gaussian approximation. However, the system relaxes into a state where densities are Gaussian distributed with mean $a_0$ and $p_0$ respectively. Therefore it is safe to naively interpret such shifts as density perturbations. Given that the bare mass term of the passive particles equals $\kappa a_0$, the initial density $a_0$ also indicates the distance away from criticality in the Gaussian approximation.

\subsection{Calculation of the structure factor}\label{subsec:Drawing}
In the Gaussian theory, observables are calculated using action~\eqref{eq:GaussianAction} which only allows for tree-level Feynman diagrams. Using the covariance in spatial Fourier space and real temporal space between active particle densities as an example, this physical observable is calculated in Doi-Peliti field theory as
\begin{equation}
\begin{split}
    \langle\rho_A\rho_A'\rangle&=\langle a^{\dagger}aa'^{\dagger}a'\rangle=\langle aa'^{\dagger}a'\rangle=\left<a\Tilde{a}'a'\right>+\left<aa'\right>\\
    &=a_0^2+a_0\left<\check a\right>+a_0\left<\check a'\right>+\left<\check a\check a'\right>+a_0\left<\check a\Tilde{a}'\right>+\left<\check a\Tilde{a}'\check a'\right>\\
\end{split}
\end{equation}
where a prime indicates that the field has arguments $(-q,t')$, whereas without the prime the field has arguments $(q,t)$. Then, 
\begin{equation}\label{eq:VarCalculationDP}
 \text{Cov}[\rho_A,\rho_A']=a_0\left<\check a\Tilde{a}'\right>+\left<\check a\check a'\right>-\left<\check a\right>\left<\check a'\right>+\left<\check a\Tilde{a}'\check a'\right>
\end{equation}

In the Gaussian approximation, the third and fourth terms give zero contributions due to the lack of source terms, such as $\Tilde{a}$, in the action. Aside from the prefactor $a_0$, the first term equals the propagator, which can be read off from the action in Fourier space
\begin{equation}
    a_0\left<\check a\Tilde{a}\right>=a_0\frac{1}{-i\omega+Dq^2}
\end{equation}
The second term $\langle aa'\rangle$ can be calculated in Fourier space as the sum of two terms represented by the following two Feynman diagrams:
\begin{equation}
\tikz[thick,baseline=-2.5pt]{\draw[color=red] (-1,-0.6) node[left]{$t'$} --(0,0) node[above right,black] {\small$\kappa p_0a_0$};\draw[color=red] (-1,0.6) node[left]{$t$} --(0,0) ;}
\hat{=}\int\frac{2\kappa p_0a_0e^{-i\omega(t-t')}}{(-i\omega+Dq^2)(i\omega+Dq^2)}\dd\omega
\end{equation}
where the factor 2 is the symmetry factor, and 
\begin{align}
\tikz[thick,baseline=-2.5pt]{\draw[color=blue,decorate,decoration=snake] (-0.5,-0.3)--(0,0) node[above right,black] {\small$-\kappa p_0a_0$};\draw[color=red] (-1,0.6) node[left]{$t$}--(0,0) ;\draw[color=red](-1,-0.6) node[left]{$t'$}--(-0.5,-0.3)}\hat=&-\kappa p_0a_0\int\frac{e^{-i\omega t-i\omega't'}\kappa a_0\delta(\omega+\omega')}{(-i\omega+Dq^2)(-i\omega'+Dq^2)(-i\omega'+\kappa a_0)}\dd\omega\dd\omega'
\end{align}
which should then be symmetrized between $\omega\longleftrightarrow\omega'$. Adding these contributions together and applying the inverse Fourier transform integral in time gives the active-active covariance function
\begin{equation}
\text{Cov}[\rho_A,\rho_A']_{\text{Gaussian}}=a_0\Biggl(e^{-Dq^2|t-t'|}+\kappa p_0\frac{Dq^2e^{-Dq^2|t-t'|}-\kappa a_0e^{-\kappa a_0|t-t'|}}{(Dq^2+\kappa a_0)(Dq^2-\kappa a_0)}\Biggr)
\end{equation}

The structure factor is an equal-time covariance, and hence we take $t=t'$ (dropping the prime on the second $\rho_A$) to obtain
\begin{equation}
    \text{Cov}[\rho_A,\rho_A]=a_0+p_0\frac{\kappa a_0}{Dq^2+\kappa a_0}
\end{equation}

Similarly the passive-passive covariance equals 
\begin{equation}\label{eq:VarCalculationDP2}
 \text{Cov}[\rho_P,\rho_P']=p_0\left<\check p\Tilde{p}'\right>+\left<\check p\check p'\right>-\left<\check p\right>\left<\check p'\right>+\left<\check p\Tilde{p}'\check p'\right>
\end{equation}
Since there is no tree-level diagram that represents a two-point passive-passive noise vertex, the only contributing part is the passive propagator multiplied by $p_0$, and hence we directly obtain
\begin{equation}
    \text{Cov}[\rho_P,\rho_P']=p_0e^{-\kappa a_0|t-t'|}
\end{equation}

Finally, the active-passive covariance has contributions from the transmutation vertex (viewed as part of the bilinear propagators) and the active-passive noise vertex. Adding all contributions together, we eventually arrive at the total density covariance
\begin{align}\label{eq:CovTotalDensity}
\text{Cov}[\rho,\rho']=a_0e^{-Dq^2|t-t'|}\Biggl(1-\frac{Dq^2\kappa p_0}{(Dq^2+\kappa a_0)(Dq^2-\kappa a_0)}\Biggr)+p_0e^{-\kappa a_0|t-t'|}\Biggl(\frac{(Dq^2)^2}{(Dq^2)^2-(\kappa a_0)^2}\Biggr)
\end{align}

Taking $t=t'$ gives the structure factor,
\begin{align}\label{eq:StructFact}
    S(q)=a_0+p_0\frac{Dq^2}{Dq^2+\kappa a_0}=a_0+p_0\frac{(q\xi)^2}{1+(q\xi)^2}
\end{align}
where $\xi=\sqrt{D/(\kappa a_0)}$ is the correlation length (confirming critical exponent $\nu=1/2$). Eq.~\eqref{eq:StructFact} also confirms the mean-field critical exponent $\eta=2$ and Eq.~\eqref{eq:CovTotalDensity} confirmed the dynamic mean-field exponent $z=2$. Here, $S(0)$ tends to 0 on approaching criticality but $S(q\neq0)$ tends to $p_0\neq0$. These results confirm Eqs.(2,3) of the main text and the resulting discontinuous hyperuniformity with `exponent' $\varsigma=0+$.

\subsection{Cole-Hopf Transformation}\label{subsec:Cole}
It is non-trivial to derive the noise in the Langevin equations corresponding to the Gaussian approximation. In fact, naively using a response-field formalism yields imaginary noise in most reaction-diffusion systems, including C-DP \cite{Stenull}. This is because, in Doi-Peliti field theory, annihilation fields are not exactly density fields. Hence, we turn to a Cole-Hopf transformation of the Doi-Peliti action \cite{Biroli} in order to obtain an action in terms of active and passive particle density fields, $\alpha$ and $\pi$, and their response fields $\Tilde\alpha$ and $\Tilde\pi$, where 
\begin{align}
    a=\exp(-\Tilde{\alpha})\alpha, &\qquad a^{\dagger}=\exp(\Tilde{\alpha})\\
    p=\exp(-\Tilde{\pi})\pi, &\qquad p^{\dagger}=\exp(\Tilde{\pi})
\end{align}
Again performing a shift $\alpha\to\alpha+\alpha_0$ and $\pi\to\pi+\pi_0$ and keeping only the quadratic parts, the Cole-Hopf action for C-DP is at Gaussian level
\begin{align}\label{eq:CH_Gauss_Action}
    \mathbb{A}_{\text{G,CH}}=-\Tilde{\alpha}\partial_t\alpha+D(-\nabla\Tilde{\alpha}\nabla\alpha+(\nabla\Tilde{\alpha}^2)\alpha_0)-\Tilde{\pi}\partial_t\pi+\kappa(\Tilde{\alpha}-\Tilde{\pi})\alpha_0\pi+\mu(\Tilde{\alpha}-\Tilde{\pi})^2\alpha_0
\end{align}

There are two advantages of this transformation. Firstly, since $\alpha$ and $\pi$ are particle density fields, fluctuations in densities can be written neatly as $\langle\alpha\alpha\rangle$, in comparison to \eqref{eq:VarCalculationDP}. Therefore calculations of correlation functions at tree-level become easier. Secondly, it is now possible to use a response-field formalism \cite{Tauber} to derive the corresponding noise in the Langevin equations: the quadratic terms in the creation fields, $D(\nabla\Tilde{\alpha})^2\alpha_0$ and $\mu(\Tilde{\alpha}-\Tilde{\pi})^2\alpha_0$, now represent noise. Starting with the second term, since
\begin{equation}
    e^{\int\dd t\dd^d x\,\mu\alpha_0(\Tilde{\alpha}-\Tilde{\pi})^2}\propto \int \dd\eta \, e^{\int\dd t\dd^d x\,(\Tilde{\alpha}-\Tilde{\pi})\eta-\eta^2/(4\mu \alpha_0)}
\end{equation}
we derive that at the level of the Gaussian approximation, there is a birth-and-death noise of variance $2\mu\alpha_0$. This noise cancels between active and passive evolution equations, and therefore does not show up in total density Langevin equation. For the first term, notice that 
\begin{equation}
    e^{\int\dd t\dd^d x \,D\alpha_0(\nabla\Tilde{\alpha})^2}\propto\int\dd\boldsymbol{\Xi} \,e^{\int\dd t\dd^d x \,\boldsymbol{\Xi}\cdot\nabla\Tilde{\alpha}-\boldsymbol{\Xi}^2/(4D\alpha_0)}
    \end{equation}
    
Using integration by parts, this term is equivalent to having a diffusional noise in the form of $\nabla\cdot\boldsymbol{\Xi}$, where $\boldsymbol{\Xi}$ is a Gaussian white noise with variance $2D\alpha_0$. The Langevin equations in the Gaussian theory then read
\begin{subequations}\label{eq:LangevinEq}
\begin{align}
    \frac{\partial \rho_A}{\partial t}&=D\nabla^2 \rho_A+\kappa \alpha_0(\rho-\rho_A-\pi_0)+\sqrt{2\mu \alpha_0}\eta+\sqrt{2D\alpha_0}\nabla\cdot\boldsymbol{\Xi}\label{eq:CH1}\\
    \frac{\partial\rho}{\partial t}&=D\nabla^2\rho_A+\sqrt{2D\alpha_0}\nabla\cdot\boldsymbol{\Xi}\label{eq:CH2}
\end{align}\end{subequations}
The density fields $\rho_A$ and $\rho_P=\rho-\rho_A$ in these Langevin equations are the same as the $\alpha$ and $\pi$ fields in the Cole-Hopf-transformed and Gaussianized action~\eqref{eq:CH_Gauss_Action}, i.e.~all their moments agree. This confirms Eqs.(4,5) of the main text.

\section{Perturbative RG Analysis}\label{sec:RG}

We follow the standard procedure for perturbative RG analysis: coarse-graining, rescaling fields and calculating 1-loop corrections for dimensions $d=d_c-\epsilon$ just below the upper critical dimension $d_c=4$. Subsequently, beta equations for effective coupling constants are solved to give the RG fixed point, which then gives rise to several independent critical exponents. 

\subsection{Rescaling Fields}\label{subsec:Rescaling}
We consider a rescaling of magnitude $\zeta$ in space coordinates $x$ and wavenumbers in Fourier space $q$, 
\begin{align}
    x &\rightarrow \frac{x}{\zeta}\\
    q &\rightarrow \zeta q
\end{align}
A shorthand representation for this rescaling can be written as $[x]=-1$, meaning length scales as inverse length scale to the power of $-1$. Typically, it should then be possible to find the scaling dimensions of the fields by dimension counting. However, rescaling of the fields is non-trivial in the C-DP model: creation and annihilation operators do not necessarily scale equally. This introduces two extra degrees of freedom that remain in question. For now, we assume that creation and annihilation fields have the same scalings at bare level. The reason for this assumption will become clearer in later discussions of the structure factor and the order parameter exponent. In this case $[\check a]=[\Tilde{a}]=[\check p]=[\Tilde{p}]=d/2$. The mass term in the passive propagator $-i\omega+\kappa a_0:=-i\omega+\epsilon_p$, determining distance from criticality, then has dimension $[\epsilon_p]=2$. The coupling constants also scale according to
\begin{align}
    [\tau_p]&=[\alpha_i]=2\\
    [\lambda_i]&=[\sigma_i]=2-\frac{d}{2}\\
    [\chi_i]&=2-d
\end{align}
Therefore simply from looking at the scaling dimensions of the coupling constants, it can be concluded that the C-DP model has an upper critical dimension of 4, above which $\chi$, $\lambda$ and $\sigma$ vertices become irrelevant, and our results for the Gaussian approximation hold. Below 4 dimensions, $\lambda$ and $\sigma$ vertices become relevant.

\subsection{Counting Loops}\label{subsec:Counting}
In $d=4-\epsilon$ dimensions, we perform perturbative renormalization in 1-loop order. Due to the relatively large number of interaction vertices, a careful enumeration of all relevant 1-loop corrections is required. Here we use Euler's theorem for a connected planar graph
\begin{equation}
    V-E+L=1 \label{eq:Euler}
\end{equation}
where $V$, $E$, $L$ represent the number of vertices, edges and loops respectively. 

Consider loops constructed by the interaction vertices $\alpha_i$, $\sigma_i$ and $\lambda_i$. We first require the number of additional outgoing legs to match the number of additional incoming legs. Noting that an $\alpha$ interaction vertex introduces two extra outgoing legs, a $\sigma$ interaction vertex introduces two extra outgoing legs and one extra incoming leg, and a $\lambda$ interaction vertex introduces one extra outgoing leg and two extra incoming legs, we obtain (in an obvious notation)
\begin{equation}\label{eq:loops}
    2\#\alpha+\#\sigma=\#\lambda
\end{equation}
One can be more precise about the number of additional active and passive propagators (diagrammatically straight, red or wavy, blue legs), but this turns out to be unnecessary for the enumeration of all possible loops. Now after these connect up to form a 1-loop correction to some interaction vertex (with vertex number $V_0$ and edge number $E_0$), the extra number of graph vertices is simply the number of interaction vertices added, or
\begin{equation}
    V=\#\alpha+\#\sigma+\#\lambda+V_0
\end{equation}

The additional interaction vertices also contribute to extra edges, but as they connect up each contribution needs to be halved, therefore
\begin{equation}
    E=\#\alpha+\frac{3}{2}\#\sigma+\frac{3}{2}\#\lambda+E_0
\end{equation}
Substituting the expressions above into \eqref{eq:Euler}, we find that $\#\sigma+\#\lambda=2$. Combined with equation \eqref{eq:loops}, this says that each 1-loop correction must either be constructed by one $\sigma$ and one $\lambda$ interaction vertex, or two $\lambda$ and one $\alpha$ vertex. There are tens of these loops, but most of their corrections cancel eventually; we will show those that will contribute in next Subsection.

\subsection{Loops contributing to corrections of $\alpha$, $\lambda$ and $\sigma$ vertices}\label{subsec:Loops}
Loops formed by one $\sigma$ and one $\lambda$ vertex that give non-zero contributions to the beta equations are shown below. The following diagrams only show the loop and not the external legs. (To see its correction to an interaction vertex simply attach external legs accordingly.) Here, each node that can be connected to external legs is marked with a black dot. As an example,we give below the loops constructed by a $\lambda$ and a $\sigma$ vertex:

\begin{center}
\tikz[thick,baseline=-2.5pt]{\draw[color=red] (-2,0) node[circle,fill=black, inner sep=1.5pt]{}--(0,0) node[circle,fill=black, inner sep=1.5pt] {} ;\draw[color=blue, decorate, decoration=snake] (-2,0) --(-1,1)  node[circle,fill=black, inner sep=1.5pt]{};\draw[color=red](-1,1) --(0,0)}\qquad\tikz[thick,baseline=-2.5pt]{\draw[color=blue, decorate, decoration=snake] (-2,0)node[circle,fill=black, inner sep=1.5pt]{}--(0,0) node[circle,fill=black, inner sep=1.5pt]{};\draw[color=red] (-2,0) --(-1,1) node[circle,fill=black, inner sep=1.5pt]{};\draw[color=red](-1,1) --(0,0)}\qquad\tikz[thick,baseline=-2.5pt]{\draw[color=red] (-2,0)node[circle,fill=black, inner sep=1.5pt]{}--(0,0)node[circle,fill=black, inner sep=1.5pt]{} ;\draw[color=blue, decorate, decoration=snake] (-2,0) --(-1,1) node[circle,fill=black, inner sep=1.5pt]{};\draw[color=blue, decorate, decoration=snake](-1,1) --(0,0)}\qquad\tikz[thick,baseline=-2.5pt]{\draw[color=blue, decorate, decoration=snake] (-2,0)node[circle,fill=black, inner sep=1.5pt]{}--(0,0) node[circle,fill=black, inner sep=1.5pt]{};\draw[color=red] (-2,0) --(-1.5,0.5);\draw[color=blue, decorate, decoration=snake] (-1.5,0.5) --(-1,1)node[circle,fill=black, inner sep=1.5pt]{} ;\draw[color=red](-1,1) --(0,0)}\qquad\tikz[thick,baseline=-2.5pt]{\draw[color=red] (-2,0)node[circle,fill=black, inner sep=1.5pt]{}--(-1,0) ;\draw[color=blue, decorate, decoration=snake] (-1,0)--(0,0)node[circle,fill=black, inner sep=1.5pt]{} ;\draw[color=blue, decorate, decoration=snake] (-2,0) --(-1,1) ;\draw[color=red](-1,1)node[circle,fill=black, inner sep=1.5pt]{} --(0,0)}\\
\end{center}

Also as an example, we now present how the third loop shown above corrects a $\sigma_3$ vertex, or \tikz[thick]{\draw[color=blue, decorate, decoration=snake] (-0.5,-0.3)--(0,0) node[above,black] {\tiny$\sigma_3$};\draw[color=red] (-0.5,0.3)--(0,0);\draw[color=red] (0,0) -- (0.7,0);}. The only way to attach this at one-loop order is shown below (since the other way to attach external outgoing legs requires the passive-passive noise vertex which has zero bare value and hence is not of 1-loop order). This is evaluated at vanishing external wavenumbers and frequencies:
\begin{equation}
\tikz[thick,baseline=9pt]{\draw[color=red] (-2,0)--(0,0);\draw[color=blue, decorate, decoration=snake] (-2,0) --(-1,1);\draw[color=blue, decorate, decoration=snake](-1,1) --(0,0);
\draw[color=red] (-1,1) --(-1.5,1.5);
\draw[color=red] (0,0) --(0.8,0);
\draw[color=blue, decorate, decoration=snake] (-2,0) --(-2.8,0);}\,
\hat{=}\,\lambda_2\sigma_3\sigma_4\int_{\boldsymbol{q}}\frac{\dd^dq}{(Dq^2+\kappa a_0)^2}
=\lambda_2\sigma_3\sigma_4\frac{(\kappa a_0)^{d/2-2}}{D^{d/2}}\frac{\Gamma(2-\frac{d}{2})}{2^d\pi^{d/2}}
\end{equation}
where in the second equality we use dimensional regularization \cite{Tauber} to deal with the divergence of loop integrals.\\

Similarly, we list next the loops constructed by two $\lambda$ vertices and one $\alpha$ vertex which have non-cancelling contributions

\begin{center}
\tikz[thick,baseline=-2.5pt]{\draw[color=blue, decorate, decoration=snake](-0.6,0)node[circle,fill=black, inner sep=1.5pt]{}--(0,0) node[circle,fill=black, inner sep=1.5pt]{}; \draw[color=blue,decorate, decoration=snake](0,0)--(0.6,0)node[circle,fill=black, inner sep=1.5pt]{};\draw[color=red](-0.6,0)--(0,0.6);\draw[color=blue,decorate,decoration=snake](0,0.6)--(1,1.6);\draw[color=red](1,1.6)--(0.6,0)}\qquad\tikz[thick,baseline=-2.5pt]{\draw[color=red](-0.6,0)node[circle,fill=black, inner sep=1.5pt]{}--(0,0) node[circle,fill=black, inner sep=1.5pt]{}; \draw[color=blue,decorate, decoration=snake](0,0)--(0.6,0)node[circle,fill=black, inner sep=1.5pt]{};\draw[color=blue,decorate,decoration=snake](-0.6,0)--(1,1.6);\draw[color=red](1,1.6)--(0.6,0)}\qquad\tikz[thick,baseline=-2.5pt]{\draw[color=red](-0.6,0)node[circle,fill=black, inner sep=1.5pt]{}--(-0.3,0);\draw[color=blue, decorate, decoration=snake](-0.3,0)--(0,0) node[circle,fill=black, inner sep=1.5pt]{}; \draw[color=blue,decorate, decoration=snake](0,0)--(0.6,0);\draw[color=blue,decorate,decoration=snake](-0.6,0)--(1,1.6);\draw[color=red](1,1.6)--(0.6,0)node[circle,fill=black, inner sep=1.5pt]{}}
\end{center}
Again as an example, we show how the the first loop corrects the $\sigma_4$ vertex, or \tikz[thick]{\draw[color=blue, decorate, decoration=snake] (-0.5,-0.3)--(0,0) node[above,black] {\tiny$\sigma_4$};\draw[color=red] (-0.5,0.3)--(0,0);\draw[color=blue,decorate,decoration=snake] (0,0) -- (0.7,0);}, with vanishing external wavenumbers and frequencies. The only way to attach this at one-loop order is as follows
\begin{multline}
\tikz[thick,baseline=9pt]{\draw[color=blue,decorate,decoration=snake](-0.6,0)--(-1.4,0);\draw[color=red](-0.7,-0.7)--(0,0);\draw[color=blue,decorate,decoration=snake](0.6,0)--(1.3,0);\draw[color=blue, decorate, decoration=snake](-0.6,0)--(0,0) ; \draw[color=blue,decorate, decoration=snake](0,0)--(0.6,0);\draw[color=red](-0.6,0)--(0,0.6);\draw[color=blue,decorate,decoration=snake](0,0.6)--(1,1.6);\draw[color=red](1,1.6)--(0.6,0)}\,
\hat{=}\,\lambda_2^2\alpha_2\sigma_4\tau_p\int_{\boldsymbol{q}}\frac{1}{2Dq^2(Dq^2+\kappa a_0)^2(\kappa a_0-Dq^2)}
+\frac{1}{4(\kappa a_0)^2(Dq^2+\kappa a_0)(Dq^2-\kappa a_0)}\\=\frac{1}{4}\frac{\lambda_2^2\alpha_2\sigma_4\tau_p}{\epsilon_p^2}\frac{(\kappa a_0)^{d/2-2}}{D^{d/2}}\frac{\Gamma(2-d/2)}{2^d\pi^{d/2}}
\end{multline}
Note that there is no contribution from this type of loop in the corrections to $\alpha_i$ and $\sigma_1$, $\sigma_3$ vertices.

\subsection{Symmetries between Coupling Constants}\label{subsec:1-PI}
At bare level, there are ten interaction vertices (Eq.~\eqref{eq:vertexDiagrams}) and ten interaction coupling constants in the Doi-Peliti action. However, many of them arise from shifts in the creation and annihilation fields and they are further linked because they originate from the gain and loss parts of a master equation. In particular, all particle reactions modelled by the master equation conserve the total particle number. (There is no extinction, coagulation nor branching present in the system.) Hence, not only do we expect that the number of relevant couplings can be reduced via relations (or symmetries) between them which must be preserved in the RG flow, but also that other couplings that emerge in the RG flow must obey these symmetries as well. There are two ways to derive the field theoretic symmetries: in the bottom-up approach we start from the master equation and trace particle number conservation through the various transformations until we reach the coupling constants. In the field-theory-intrinsic approach, we consider the loop expansions of all couplings and try to spot which symmetries are maintained in the RG flow. We outline both approaches in the following, starting with the former.

In the second-quantized version of the master equation, the key observations are that (i) gain and loss terms contain equal numbers of annihilation operators and (ii) that the loss component will always contain an equal number of creation and annihilation operators while the gain will have equal numbers of them if and only if particle numbers are conserved. (For coagulation or extinction processes, the gain term contains fewer creation than annihilation operators, whereas for branching processes, the gain term contains more creation than annihilation operators.) Furthermore, if the particle-number-conserving reaction contains a single change in particle type (single active to passive or vice versa), as is the case for our C-DP action, then loss and gain terms differ only in one creation operator while other potentially appearing creation operators are the same in both terms. This implies that the action must only contain terms that together have the factor $p^\dagger-a^\dagger$. This requirement remains true in the field theory as operators are essentially replaced by their corresponding fields.

The factor $p^\dagger-a^\dagger$ is present in the bare action and can be seen by undoing the Doi-shift. In the RG flow, new couplings that appear must still obey this symmetry, i.e.~in combination with other couplings they must maintain the factor $p^\dagger-a^\dagger$ in the interaction couplings. The couplings that group together in this way attach to vertices that have identical annihilation field legs (number and types) and the same number but not necessarily same type of Doi-shifted creation field legs. Since our theory only has two particle types, for vertices that only have a single creation field leg, coupling groups are pairs and maintaining this factor becomes straightforward, implying $\lambda_1=-\lambda_2$ and $\epsilon_p=\tau_p$ throughout  the RG flow. Similarly straightforward is the symmetry between the other, emergent couplings with one creation-field leg:
\begin{subequations}\label{eq:all_symmetries}\label{eq:all_symmetries_A}\begin{align}
\lambda_1=&\,-\lambda_2\\
\lambda_3=&\,-\lambda_4\\
\lambda_5=&\,-\lambda_6\\
\epsilon_p=&\,\tau_p\\
\epsilon_a=&\,\tau_a\\
\gamma_1=&\,-\gamma_2
\end{align} \end{subequations}
However, vertices that contain more than one creation-field leg will not have such a straightforward symmetry. In the RG flow, $\alpha_3$ for example emerges and is grouped with $\alpha_1$ and $\alpha_2$ since they have all zero annihilation field legs and two creation field legs. The coupling $\alpha_3$ could undo the $(p^\dagger-a^\dagger)$ factor between the bare $\alpha_1$ and $\alpha_2$. Undoing the Doi-shift and factorizing $p^\dagger-a^\dagger$  gives\begin{align}\label{eq:Symmetry}
\alpha_1\Tilde a^2&\,+\alpha_2\Tilde a\Tilde p+\alpha_3\Tilde p^2=\\
=&\,\alpha_1(a^\dagger-1)^2+\alpha_2(a^\dagger-1)(p^\dagger-1)+\alpha_3(p^\dagger-1)^2\notag\\
=&\,(p^\dagger-a^\dagger)(\alpha_3p^\dagger-\alpha_1a^\dagger-2\alpha_3+2\alpha_1)+a^\dagger p^\dagger(\alpha_1+\alpha_2+\alpha_3)+p^\dagger(-2\alpha_1-\alpha_2)+a^\dagger(-2\alpha_3-\alpha_2)+(\alpha_1+\alpha_2+\alpha_3)\notag
\end{align}
The final (operator-free) term ,and the $a^\dagger p^\dagger$ term, already require the symmetry $\alpha_1+\alpha_2+\alpha_3=0$. Once this relation is inserted, the $p^\dagger$ and $a^\dagger$ terms factorize as well:
\begin{align}
p^\dagger(-2\alpha_1-\alpha_2)+a^\dagger(-2\alpha_3-\alpha_2)\Bigr|_{\alpha_1=-(\alpha_2+\alpha_3)}=(p^\dagger-a^\dagger)(2\alpha_3+\alpha_2)
\end{align}
In conclusion, the required factor can only be maintained if and only if the relation $\alpha_1=-(\alpha_2+\alpha_3)$ is preserved in the RG flow. An alternative derivation is to look at the Doi-shifted creation fields: the factor $(p^{\dagger}-a^{\dagger})$ is Doi-shifted into $(\Tilde{p}-\Tilde{a})$. In the $\alpha$-vertices example in equation \eqref{eq:Symmetry}, the quadratic form has a factor of $(\Tilde{p}-\Tilde{a})$ if and only if $\alpha_1+\alpha_2+\alpha_3=0$.

Similar relations are true for $\sigma$ vertices and $\chi$ vertices (although the latter are irrelevant):
\begin{subequations}\label{eq:all_symmetries_B}\begin{align}
\alpha_1=&\,-(\alpha_2+\alpha_3)\\
\sigma_1=&\,-(\sigma_3+\sigma_5)\\
\sigma_2=&\,-(\sigma_4+\sigma_6)\\
\chi_1=&\,-(\chi_2+\chi_3)
\end{align} \end{subequations}
Finally, we express the action density $\mathbb{A}$ in terms of the \textit{running couplings} on the right hand side of Eq.~\eqref{eq:all_symmetries_A} and~\eqref{eq:all_symmetries_B}, {\em i.e.}, we make use of the symmetries, and we see that all interactions appear with a factor $p^\dagger-a^\dagger$:\begin{multline}\label{eq:shiftedAction-NoDoi}
    \mathbb{A}=-\Tilde{a}(\partial_t-D\nabla^2)\check a-\Tilde{p}\partial_t\check p+(p^\dagger-a^\dagger)\Bigl(\Tilde a(\alpha_2+\sigma_3\check a+\sigma_4\check p+\chi_2\check a\check p)+(\Tilde p+\Tilde a)(\alpha_3+\sigma_5\check a+\sigma_6\check p+\chi_3\check a\check p)\\
    +\gamma_2-\tau_p\check p-\tau_a\check a+\lambda_2\check a\check p+\lambda_4\check a\check a+\lambda_6\check p\check p\Bigr) 
   \end{multline}
If any of the symmetries~\eqref{eq:all_symmetries_A} and \eqref{eq:all_symmetries_B} were violated, the factor $p^\dagger-a^\dagger$ in the nonlinear terms would not emerge.

The above line of argument uses a microscopic conservation law on total particle number to constrain the nonlinearities to all orders. However, \eqref{eq:all_symmetries_A} and \eqref{eq:all_symmetries_B} can alternatively be found entirely at field-theoretic level by analyzing the loop expansions of the couplings. In expectation of these symmetries, we only confirm them by this route to one-loop order, shown below with the example of $\alpha_1=-(\alpha_2+\alpha_3)$. In this context it is important to be mindful of what constitutes a 1-particle irreducible (1-PI) loop diagram that should be considered as part of the corrections to an interaction vertex. The common interpretation of 1-PI loop diagrams is that we do not consider any further corrections outside the loop, including further loops and transmutations; this is because usually such corrections either are higher order in $\epsilon$ or are already generated in tree-level expansions. However, under scrutiny we find a number of loops corrections which would not appear in a tree-level expansion but are still of order $\mathcal{O}(\epsilon)$. An example is \tikz[thick,baseline=-2.5pt]{\draw[color=blue, decorate, decoration=snake] (-1.5,0) --(-1,0);\draw[color=red] (-1,1)--(-0.7,0.7) ;\draw[color=blue,decorate,decoration=snake](-0.7,0.7)--(-0.5,0.5);
\draw[color=blue,decorate,decoration=snake] (-1,0)--(0,0) ;\draw[color=red] (-1,0) --(-0.5,0.5) ;\draw[color=red](-0.5,0.5) --(0,0)}. This would not appear in usual tree-level corrections because this corresponds to the tree-level equivalent correction of \tikz[thick, baseline=-2.5pt]{\draw[color=red](-0.6,0.4)--(-0.3,0.2);\draw[color=blue,decorate,decoration=snake](-0.3,0.2)--(0,0);\draw[color=blue,decorate,decoration=snake](-0.6,-0.4)--(0,0){}}, which includes a vertex \tikz[thick, baseline=-2.5pt]{\draw[color=blue,decorate,decoration=snake](-0.6,0.4)--(0,0);\draw[color=blue,decorate,decoration=snake](-0.6,-0.4)--(0,0)node[above right]{\tiny$\alpha_3$}} that has zero bare value. Therefore, though this has a transmutation outside the loop, it would still be a 1-PI diagram, i.e.~it is necessary to include the correction of the $\alpha_3$ vertex as part of the loop corrections to $\alpha_2$. While the tree-level transmutation introduces a factor of $\frac{\epsilon_p}{-i\omega+\epsilon_p}$, at vanishing frequencies this tends to unity and hence the effective coupling constant is now $\alpha_2+\alpha_3$. Note that although we can similarly draw a correction such as \tikz[thick, baseline=-2.5pt]{\draw[color=red](-0.6,0.4)--(-0.3,0.2);\draw[color=blue,decorate,decoration=snake](-0.3,0.2)--(0,0);\draw[color=red] (-0.6,-0.4)--(-0.3,-0.2);\draw[color=blue,decorate,decoration=snake](-0.3,-0.2)--(0,0){}}, this would already be considered as part of the effective coupling constant $\alpha_2+\alpha_3$, and therefore would not be considered additionally here. 

The above conclusions at one loop are confirmed by the microscopic symmetry arguments given above which establish that $\alpha_1=-(\alpha_2+\alpha_3)$ throughout the RG flow. However, the loop analysis just presented adds something more: we can now identify the 1-loop effective coupling constants as $\alpha_1$ and $\alpha_2+\alpha_3$, and not some other combination (such as $\alpha_2$ and $\alpha_1+\alpha_3$). Doing so, we now calculate vertex functions in terms of $\alpha_1$ and $\alpha_2+\alpha_3$
\begin{align}
\Gamma_{\alpha_1}&=\alpha_1+\underline{2}\lambda_1\sigma_1\alpha_2\frac{(\kappa a_0)^{d/2-2}}{D^{d/2}}\frac{\Gamma(2-\frac{d}{2})}{2^d\pi^{d/2}}+\hdots\\
\begin{split}
\Gamma_{\alpha_2+\alpha_3}&=\alpha_2+\underline{2}\lambda_2\sigma_1\alpha_2\frac{(\kappa a_0)^{d/2-2}}{D^{d/2}}\frac{\Gamma(2-\frac{d}{2})}{2^d\pi^{d/2}}+\lambda_1\sigma_3\alpha_2\frac{(\kappa a_0)^{d/2-2}}{D^{d/2}}\frac{\Gamma(2-\frac{d}{2})}{2^d\pi^{d/2}}+\lambda_2\sigma_3\alpha_2\frac{(\kappa a_0)^{d/2-2}}{D^{d/2}}\frac{\Gamma(2-\frac{d}{2})}{2^d\pi^{d/2}}+\hdots\\
&=\alpha_2+\underline{2}\lambda_2\sigma_1\alpha_2\frac{(\kappa a_0)^{d/2-2}}{D^{d/2}}\frac{\Gamma(2-\frac{d}{2})}{2^d\pi^{d/2}}+\hdots
\end{split}
\end{align}

Here, the prefactors of 2 (underlined) arise from the symmetry factor 2 introduced by the vertex $\sigma_1$. The dots indicate that there are further terms, including of one loop order,  in the perturbative expansion. In the second line, we note that interaction vertices $\lambda_1$ and $\lambda_2$ remain equal and opposite throughout the RG flow since they have the exact same loop corrections: for each loop correction contributing to $\lambda_1$, we can replace the external outgoing straight leg with a wavy one, which represents a one-to-one correspondence to a loop correction contributing to $\lambda_2$. We then see that the corrections to the vertex functions of $\alpha_1$ and $\alpha_2+\alpha_3$ are exactly opposite, and since the two quantities have opposite values at bare level, they will likewise remain opposite in the RG flow. 

Using exactly the same field-theoretic arguments, we can deduce similar symmetries for the $\sigma$ vertices, namely between $\sigma_1$ and $\sigma_3+\sigma_5$, and between $\sigma_2$ and $\sigma_4+\sigma_6$. 
Such considerations do not affect loop corrections to any further quantities. In particular, the above arguments seem to give extra loop corrections to the $\lambda$ vertices, represented diagrammatically as \tikz[thick,baseline=-2.5pt]{;\draw[color=red] (0,1)--(-0.5,0.5) ;
\draw[color=red] (0,0) --(0.3,0) ;
\draw[color=blue, decorate, decoration=snake] (0.3,0) --(0.6,0) ;
\draw[color=red] (0,0) --(-1,0) node[circle,fill=black, inner sep=1.5pt]{};\draw[color=blue, decorate, decoration=snake] (-1,0) --(-0.5,0.5) ;\draw[color=blue, decorate, decoration=snake](-0.5,0.5) --(0,0)} (again the node marked is connected to external legs): this is because while the equivalent tree-level diagram \tikz[thick,baseline=-2.5pt]{\draw[color=red] (0.6,0.36)--(0,0) node[circle,fill=black,inner sep=1.5pt]{} ;\draw[color=red](0,0)--(0.3,-0.18);\draw[color=blue,decorate,decoration=snake](0.3,-0.18)--(0.6,-0.36)} is absent, the loop itself is still of order $\mathcal{O}(\epsilon)$ and should be included. However, we note that this loop correction is proportional to a factor of $\frac{\epsilon_p}{-i\omega+Dq^2}$. In the scaling regime where $q\xi\gg 1$, its contribution vanishes and therefore does not change the beta equations for $\lambda$ vertices.

{\em Note:} We thank Marius Bothe and Gunnar Pruessner for valuable discussions and insights that eventually led to the results presented in this subsection.

\subsection{Beta Equations}\label{subsec:Beta}
Following standard procedures \cite{Tauber}, we first define the $Z$-factors indicating RG flow of fields and coupling constants
\begin{align}
    &D_R=Z_DD, \check a_R=Z_{\check a}^{1/2}\check a,  \Tilde{a}_R=Z_{\Tilde{a}}^{1/2}\Tilde{a}, \check p_R=Z_{\check p}^{1/2}\check p,  \Tilde{p}_R=Z_{\Tilde{p}}^{1/2}\Tilde{p}\\ 
    &\alpha_R=Z_{\alpha}\alpha\zeta^{-2},  \lambda_R=A_d^{1/2}Z_{\lambda}\lambda\zeta^{(d-4)/2}, \sigma_R=A_d^{1/2}Z_{\sigma}\sigma\zeta^{(d-4)/2}\\
   \text{with } &A_d:=\frac{\Gamma(3-d/2)}{2^{d-1}\pi^{d/2}} \text{ (a constant arising from angular integrals)}
\end{align}
Then define the vertex functions where there are $k$ outgoing active propagators, $l$ incoming active propagators, $m$ outgoing passive propagators, $n$ incoming passive propagators to be $\Gamma^{(k,l,m,n)}$. We start with the field renormalizations. The active propagator is corrected by the following loop diagrams:
\begin{align}
\tikz[thick,baseline=-2.5pt]{\draw[color=red](-1,0)--(0,0)}_R\,\hat{=}\,
\tikz[thick,baseline=-2.5pt]{\draw[color=red](-1,0)--(0,0)}
+
\tikz[thick,baseline=-2.5pt]{\draw[color=red](-1,0)--(0,0); \draw[color=red](0,0)to[out=90,in=90](1,0);\draw[color=blue,decorate,decoration=snake](0,0)to[out=270,in=270](1,0);\draw[color=red](1,0)to(2,0)}
+
\tikz[thick,baseline=-2.5pt]{\draw[color=red](-1,0)--(0,0);\draw[color=blue,decorate,decoration=snake](0,0)--(1,0);\draw[color=red](1,0)--(2,0);\draw[color=red](0,0)--(1.5,1);\draw[color=blue,decorate,decoration=snake](1,0)--(1.5,1)}
\end{align}
This corresponds to the corrections to vertex functions
\begin{multline}
    \Gamma^{(1,1,0,0)}(q,\omega)=i\omega+Dq^2+\lambda_1\sigma_3\int_{\boldsymbol{k}}\frac{1}{Dk^2+\kappa a_0+i\omega}+\lambda_1\lambda_2\alpha_2\int_{\boldsymbol{k}}\frac{1}{(D(q-k)^2+\kappa a_0)(D(q-k)^2+\kappa a_0+i\omega)}
\end{multline}

Performing a shift by $q$ in the second integral reveals the fact that neither of the corrections is $q$-dependent. Similarly, the passive propagator is corrected by:\begin{align}
\tikz[thick,baseline=-2.5pt]{\draw[color=blue,decorate,decoration=snake](-1,0)--(0,0)}_R\,\hat{=}\,&
\tikz[thick,baseline=-2.5pt]{\draw[color=blue,decorate,decoration=snake](-1,0)--(0,0)}
+
\tikz[thick,baseline=-2.5pt]{\draw[color=blue,decorate,decoration=snake](-1,0)--(0,0); \draw[color=red](0,0)to[out=90,in=90](1,0);\draw[color=blue,decorate,decoration=snake](0,0)to[out=270,in=270](1,0);\draw[color=blue,decorate,decoration=snake](1,0)to(2,0)}
+
\tikz[thick,baseline=-2.5pt]{\draw[color=blue,decorate,decoration=snake](-1,0)--(0,0);\draw[color=blue,decorate,decoration=snake](0,0)--(1,0);\draw[color=blue,decorate,decoration=snake](1,0)--(2,0);\draw[color=red](0,0)--(1.5,1);\draw[color=red](1,0)--(1.5,1)}
+
\tikz[thick,baseline=-2.5pt]{\draw[color=blue,decorate,decoration=snake](-1,0)--(0,0);\draw[color=red](0,0)--(1,0);\draw[color=blue,decorate,decoration=snake](1,0)--(2,0);\draw[color=blue,decorate,decoration=snake](0,0)--(1.5,1);\draw[color=red](1,0)--(1.5,1)}
    +
\tikz[thick,baseline=-2.5pt]{\draw[color=blue,decorate,decoration=snake](-1,0)--(0,0);\draw[color=blue,decorate,decoration=snake](0,0)--(1,0);\draw[color=blue,decorate,decoration=snake](1,0)--(2,0);\draw[color=red](0,0)--(1.5,1);\draw[color=red](1,0)--(1.25,0.5);\draw[color=blue,decorate,decoration=snake](1.25,0.5)--(1.5,1)}\\
    &+
\tikz[thick,baseline=-2.5pt]{\draw[color=blue,decorate,decoration=snake](-1,0)--(0,0);\draw[color=red](0,0)--(0.5,0);\draw[color=blue,decorate,decoration=snake](0.5,0)--(1,0);\draw[color=blue,decorate,decoration=snake](1,0)--(2,0);\draw[color=blue,decorate,decoration=snake](0,0)--(1.5,1);\draw[color=red](1,0)--(1.5,1)}
    +
 \tikz[thick,baseline=-2.5pt]{\draw[color=blue,decorate,decoration=snake](-1,0)--(0,0);\draw[color=blue,decorate,decoration=snake](0,0)--(1,0);\draw[color=blue,decorate,decoration=snake](1,0)--(2,0);\draw[color=red](0,0)--(1,0.666);\draw[color=blue,decorate,decoration=snake](1,0.666)--(1.5,1);\draw[color=red](1,0)--(1.5,1)}
\end{align}
corresponding to
\begin{multline}
    \Gamma^{(0,0,1,1)}(q,\omega)=i\omega+\kappa a_0+\lambda_2\sigma_4\int_{\boldsymbol{k}}\frac{1}{Dk^2+\kappa a_0+i\omega}+2\lambda_2^2\alpha_1\int_{\boldsymbol{k}}\frac{1}{(Dk^2+\kappa a_0+i\omega)(2Dk^2)}\\
    +\lambda_1\lambda_2\alpha_2\int_{\boldsymbol{k}}\frac{1}{(Dk^2+\kappa a_0)(i\omega+D(q+k)^2+\kappa a_0)}+
    \lambda_2^2\tau_p\alpha_2\int_{\boldsymbol{k}}\frac{1}{2Dk^2(Dk^2+\kappa a_0)(Dk^2+\kappa a_0+i\omega)}\\
    +\lambda_2^2\tau_p\alpha_2\int_{\boldsymbol{k}}\frac{1}{(Dk^2+\kappa a_0)(i\omega+2\kappa a_0)(D(q+k)^2+\kappa a_0+i\omega)}\\
    +\lambda_2^2\tau_p\alpha_2\int_{\boldsymbol{k}}\frac{1}{(\kappa a_0-Dk^2)(2Dk^2)(Dk^2+\kappa a_0+i\omega)}+\frac{1}{(Dk^2-\kappa a_0)(Dk^2+\kappa a_0)(i\omega+2\kappa a_0)}
\end{multline}
The $Z$-factors for the fields are given by
\begin{align}
Z_{\check a}^{1/2}Z_{\Tilde{a}}^{1/2}&=\frac{\partial\Gamma^{(1,1,0,0)}(q,\omega)}{\partial i\omega}=1-\frac{\lambda_1\sigma_3}{D^{d/2}}\frac{A_d\zeta^{-\epsilon}}{\epsilon}\\
Z_{\check p}^{1/2}Z_{\Tilde{p}}^{1/2}&=\frac{\partial\Gamma^{(0,0,1,1)}(q,\omega)}{\partial i\omega}=1-\frac{\lambda_2\sigma_4}{D^{d/2}}\frac{A_d\zeta^{-\epsilon}}{\epsilon}-\frac{1}{2}\frac{\lambda_2^2\alpha_2\tau_p}{\epsilon_p^2}\frac{1}{D^{d/2}}\frac{A_d\zeta^{-\epsilon}}{\epsilon}
\end{align}
Then from $\partial\Gamma^{(1,1,0,0)}(q,\omega)/\partial q^2$ and $\partial \Gamma^{(0,0,1,1)}(q,\omega)/\partial \epsilon_p$ we obtain
\begin{align}
    Z_{\check a}^{1/2}Z_{\Tilde{a}}^{1/2}Z_D&=1\\
    Z_{\check p}^{1/2}Z_{\Tilde{p}}^{1/2}Z_{\epsilon_p}&=1-\frac{\lambda_2\sigma_4}{D^{d/2}}\frac{A_d\zeta^{-\epsilon}}{\epsilon}-\frac{\lambda_2^2\alpha_2\tau_p}{\epsilon_p^2}\frac{1}{D^{d/2}}\frac{A_d\zeta^{-\epsilon}}{\epsilon}
\end{align}
Hence, we have found the $Z$-factors for the diffusion constant $D$, $Z_D=1+\frac{\lambda_1\sigma_3}{D^{d/2}}\frac{A_d\zeta^{-\epsilon}}{\epsilon}$, and the true distance to the critical point $\epsilon_p$, $Z_{\epsilon_p}=1-\frac{\lambda_2^2\alpha_2\tau_p}{2\epsilon_p^2D^{d/2}}\frac{A_d\zeta^{-\epsilon}}{\epsilon}$. The scalings of these give rise to the dynamic exponent $z$ and correlation length exponent $\nu_{\perp}$ respectively.\\

We identify the effective coupling constants
\begin{align}
    u_R&=\frac{\lambda_1\sigma_3}{D^{d/2}}A_d\zeta^{-\epsilon}Z_{\lambda_1}Z_{\sigma_3}\label{eq:EffectiveBeta1}\\
    v_R&=\frac{\lambda_2\sigma_4}{D^{d/2}}A_d\zeta^{-\epsilon}Z_{\lambda_2}Z_{\sigma_4}\label{eq:EffectiveBeta2}\\
    w_R&=\frac{\lambda_2^2\alpha_2\tau_p}{\epsilon_p^2D^{d/2}}A_d\zeta^{-\epsilon}\frac{Z_{\lambda_2}^2Z_{\alpha_2}}{Z_{\epsilon_p}}\label{eq:EffectiveBeta3}
\end{align}
RG fixed points are found by setting the beta equations for the effective coupling constants \eqref{eq:EffectiveBeta1}-\eqref{eq:EffectiveBeta3} equal to zero. 

To demonstrate this, we use $u_R$ as an example. The $\lambda_1$ vertex is corrected by all the five $[\lambda\sigma]$ loops and all the three $[\lambda\lambda\alpha]$ loops, giving the following relation
\begin{equation}
Z_{\lambda_1}Z_{\Tilde{a}}^{1/2}Z_{\check a}^{1/2}Z_{\check p}^{1/2}=1+\frac{\lambda_1\sigma_3}{D^{d/2}}\frac{A_d\zeta^{-\epsilon}}{\epsilon}+\frac{\lambda_2\sigma_4}{D^{d/2}}\frac{A_d\zeta^{-\epsilon}}{\epsilon}
\end{equation}
Similarly, the $\sigma_3$ vertex is corrected by all the five $[\lambda\sigma]$ loops and none of the three $[\lambda\lambda\alpha]$ loops, therefore giving
\begin{equation}
Z_{\sigma_3}Z_{\Tilde{a}}^{1/2}Z_{\Tilde{p}}^{1/2}Z_{\check a}^{1/2}=1+2\frac{\lambda_1\sigma_3}{D^{d/2}}\frac{A_d\zeta^{-\epsilon}}{\epsilon}+\frac{\lambda_2\sigma_4}{D^{d/2}}\frac{A_d\zeta^{-\epsilon}}{\epsilon}
\end{equation}
From these $Z$-factors we obtain the flow functions 
\begin{align}
\left.\gamma_{\lambda_1}=\zeta\frac{\partial}{\partial\zeta}\right\vert_0\ln Z_{\lambda_1}&=-2+\frac{d}{2}-u_R-v_R-\frac{1}{2}(\gamma_{\Tilde{a}}+\gamma_{a}+\gamma_{p})\\
\left.\gamma_{\sigma_3}=\zeta\frac{\partial}{\partial\zeta}\right\vert_0\ln Z_{\sigma_3}&=-2+\frac{d}{2}-2u_R-v_R-\frac{1}{2}(\gamma_{\Tilde{a}}+\gamma_{\Tilde{p}}+\gamma_{a})\\
\left.\gamma_{D}=\zeta\frac{\partial}{\partial\zeta}\right\vert_0\ln Z_{D}&=-u_R
\end{align}
Therefore the renormalization group beta function for the effective coupling constant $u_R$ reads
\begin{equation}
    \beta_u=\left.\zeta\frac{\partial}{\partial\zeta}\right\vert_0u_R=u_R(\gamma_{\lambda_1}+\gamma_{\sigma_3}-\frac{d}{2}\gamma_D)=u_R\left[-\epsilon-3u_R-3v_R-\frac{1}{2}w_R+\mathcal{O}(\epsilon^2)\right]
\end{equation}
Notice that there is no need to settle degrees of freedom in field renormalization at this point, since they always show up in pairs of $\left(\Tilde{a},\check a\right)$ and $\left(\Tilde{p},\check p\right)$ in the beta functions of effective coupling constants. Applying similar methods to the remaining couplings, we have
\begin{align}
    \gamma_{\sigma_4}&=-2+\frac{d}{2}-2u_R-v_R+\frac{w_R}{2}-\frac{1}{2}(\gamma_{\Tilde{a}}+\gamma_{\Tilde{p}}+\gamma_{p})\\
    \gamma_{\alpha_2}&=-2-2u_R-v_R-\frac{1}{2}(\gamma_{\Tilde{a}}+\gamma_{\Tilde{p}})
\end{align}
Substituting these in \eqref{eq:EffectiveBeta1}-\eqref{eq:EffectiveBeta3} we find
\begin{align}
    3u_R+3v_R+\frac{1}{2}w_R&=-\epsilon\\
    2u_R+4v_R+\frac{1}{2}w_R&=-\epsilon\\
    4u_R+5v_R+\frac{3}{2}w_R&=-\epsilon
\end{align}
Solving these linear equations gives $u_R=v_R=-2\epsilon/9$ and $w_R=2\epsilon/3$. These results tell us the scaling dimensions for the diffusion constant $D$ and order parameter $\epsilon_p$, from which we can derive the dynamic exponent $z=2+u_R=2-2\epsilon/9$ and the correlation length exponent $1/\nu_{\perp}=2-1/2w_R=2-\epsilon/3$. These match to order $\epsilon$ the results found via the mapping onto the quenched Edwards-Wilkinson model (q-EW) via functional RG \cite{2loopFRGqEW}. Since the beta functions are linear and of full rank, there is only one fixed point.

There are two further things to note. Firstly, these results give us the sum of anomalous dimensions of the annihilation and creation of fields $[\Tilde{a}\check a]=d+2\epsilon/9$, $[\Tilde{p}\check p]=d-\epsilon/9$. They do not give the dimensions of $\Tilde{a},\check a, \Tilde{p}, \check p$ separately; these will be discussed in the next Section. Secondly, the shifts of the annihilation fields ($a_0$ and $p_0$) cannot be expressed as any combination of the effective coupling constants $u_R$, $v_R$ and $w_R$. Since all universal behaviour to order $\mathcal{O}(\epsilon)$ should be defined by these RG fixed point values, we deduce that these shifts are non-universal. This is not surprising: as previously explained, they describe initialization of the system in the far past.

\section{Cancellations in the Structure Factor}\label{sec:Cancellations}
The structure factor for the total density consists of three parts: the active covariance \eqref{eq:VarCalculationDP}, the passive covariance \eqref{eq:VarCalculationDP2} and the active-passive covariance. The active-passive covariance is calculated as follows 
\begin{equation}
\text{Cov}[\rho_A,\rho_P']=\Theta(t-t')\left(\langle\check a\Tilde{p}'\check p'\rangle+p_0\langle\check a\Tilde p'\rangle\right)+\Theta(t'-t)\left(\langle\check p'\Tilde{a} \check a\rangle+a_0\langle\check p'\Tilde{a}\rangle\right)+\langle\check a\check p'\rangle-\langle\check a\rangle\langle\check p'\rangle
\end{equation}
While the complete form for the structure factor is lengthy, we can make the following observations:
\begin{itemize}
    \item Terms that involve transmutations, such as $\langle \check a\Tilde{p}\rangle$, vanish in the equal-time limit. Physically this is because the insertion of an active/passive density cannot instantaneously affect the density of the other species.
    \item Observables that involve three field operators such as $\langle \check a\Tilde{a}\check a\rangle$ also vanish. This is because diagrams that contribute to these are disconnected, which, though allowed in Doi-Peliti field theory, give contributions that contain pre-factors of negative powers of $\xi$ hence giving zero contributions.
    \item Terms in the form of $\langle \check a\rangle\langle \check a\rangle$ cancel eventually due to conservation of total density.
\end{itemize}

Hence, the contributing parts can be gathered to give
\begin{equation}\label{eq:StructureFactor}
S(q)=a_0\left<\check a\Tilde{a}\right>+\left<\check a\check a\right>+\left<\check a\check p\right>+\left<\check p\check a\right>+\langle \check p\check p\rangle+p_0\left<\check p\Tilde{p}\right>
\end{equation}
In the Gaussian approximation, every term in \eqref{eq:StructureFactor} has a $q^0F_i(q\xi)$ scaling behaviour. Amplitudes of all terms except for the first cancel exactly at low $q$, and hence approaching criticality $S(0)=a_0$ tends to 0. Hyperuniformity emerges in a discontinuous sense here, with the hyperuniformity exponent $\varsigma=0^+$. 

Below 4 dimensions, we expect a broadly similar cancellation to occur for all terms except the first in \eqref{eq:StructureFactor}. Note that RG already tells us that the final term should scale as $q^{-\epsilon/9}$. In addition, conservation of total particle density implies that active and passive fluctuations must scale alike, {\em i.e.}, the anomalous dimensions for active and passive annihilation fields must match: $[\check a]=[\check p]$. Hence the terms $S_{2,3,4,5}$ should all have the same scaling behaviour. Using the emergence of hyperuniformity at criticality as a constraint, it is necessary for these terms to cancel with the final $q$-divergent term. This means their scalings are the same as the final $\langle \check p\Tilde{p}\rangle$ term, implying $[\check a]=[\check p]=[\Tilde{p}]$, and that their summed amplitudes cancel in the scaling limit $q\xi\gg 1$. We can therefore use the field renormalization results $[\check a\Tilde{a}]=d+2\epsilon/9$ and $[\check p\Tilde{p}]=d-\epsilon/9$ to derive
\begin{align}
    [\check a] = [\check p] = [\Tilde{p}]=\frac{d}{2}-\frac{\epsilon}{18}\qquad [\Tilde{a}]=\frac{d}{2}+\frac{5\epsilon}{18}
\end{align}
Then using the following scaling relation \cite{Lubeck,2loopFRGqEW}
\begin{equation}
    \beta=\nu_{\perp}(d/2+\eta_{\check a})
\end{equation}
we retrieve the order parameter exponent to be $\beta=1-\epsilon/9$, matching with results found in the q-EW mapping \cite{2loopFRGqEW}. 

Since each term in \eqref{eq:StructureFactor} has a pure scaling behaviour, we can now rewrite the structure factor in the following form
\begin{equation}
    S(q) = q^{2\epsilon/9}F_1(q\xi)+q^{-\epsilon/9}\sum_{i=2}^6 F_i(q\xi):= q^{2\epsilon/9}F_1(q\xi)+q^{-\epsilon/9}G(q\xi) \label{eq:Amp}
\end{equation}
The requirement that the system is hyperuniform at criticality ($q\xi\gg$1) implies that $G(\infty)=0$. Further, since away from criticality ($\xi<\infty$) $S(0)$ must be finite, $G(s)$ needs to vanish at small $s$. If a power law (as expected by general scaling arguments), it must vanish faster than $s^{\epsilon/9}$, so as to cancel out the divergence at small $q$. However, this leaves a positive power of $\xi$, which implies a diverging amplitude, on approach to criticality, of $S(0<q\ll \xi^{-1})$. Therefore, we conclude that for the emergence of hyperuniformity, $G(s)$ must be strictly zero at small but finite $s$. The simplest way this could happen is for $G$ to vanish everywhere, as would arise by exact cancellation of the five amplitude functions in \eqref{eq:Amp}. In any case, assuming hyperuniformity does emerge at criticality, as is known to be the case for C-DP, its exponent $\varsigma$ is found from the scaling of the first term, {\em i.e.}, $\varsigma=\frac{2\epsilon}{9}+\mathcal{O}(\epsilon^2)$.  This completes the argument, outlined in the paragraphs following Eq.6 of the main text, for this value of the hyperuniformity exponent.